\documentclass[aps,prb, twocolumn,floatfix,notitlepage, superscriptaddress,longbibliography]{revtex4-2}
\usepackage[utf8]{inputenc}
\usepackage{amsfonts,amsmath,amssymb,graphicx,epstopdf,verbatim}
\usepackage{rotating}
\usepackage{epstopdf}
\usepackage{xcolor}
\usepackage{natbib}
\usepackage[colorlinks]{hyperref}
\definecolor{darkred}{rgb}{0.5,0,0}
\definecolor{darkgreen}{rgb}{0,0.5,0}
\definecolor{darkblue}{rgb}{0,0,0.5}
\hypersetup{colorlinks,
linkcolor=darkred,
filecolor=darkgreen,
urlcolor=darkblue,
citecolor=darkgreen}
\usepackage{braket}
\usepackage{overpic}

\renewcommand{\Im}[1]{\mathcal{I}\mathrm{m}\left[#1\right]}
\renewcommand{\Re}[1]{\mathcal{R}\mathrm{e}\left[#1\right]}

\newcommand{\ud}{\mathrm{d}}
\newcommand{\nep}{\operatorname{e}}

\newcommand{\opcdag}[1]{{\hat{c}^{\dagger}}_{#1}}

\newcommand{\opc}[1]{{\hat{c}^{\phantom \dagger}}_{#1}}

\newcommand{\opgammadag}[1]{{\hat{\gamma}^{\dagger}}_{#1}}
\newcommand{\opgamma}[1]{{\hat{\gamma}^{\phantom \dagger}}_{#1}}

\newcommand{\U}{{\textbf U}}

\newcommand{\Q}{{\textbf Q}}
\newcommand{\opbfPsi}[1]{{\widehat{\mathbf{\Psi}}^{\phantom \dagger}}_{#1}}
\newcommand{\opbfPsidag}[1]{{\widehat{\mathbf{\Psi}}^{\dagger}}_{#1}}
\newcommand{\opbfc}[1]{{\hat{{\mathbf c}}^{\phantom \dagger}}_{#1}}
\newcommand{\opbfcdag}[1]{{\hat{{\mathbf c}}^{\dagger}}_{#1}}
\newcommand{\B}{{\textbf B}}
\newcommand{\V}{{\textbf V}}
\newcommand{\opbfgamma}[1]{{\hat{\mbox{\boldmath $\gamma$}}^{\phantom \dagger}}_{#1}}
\newcommand{\opbfgammadag}[1]{{\hat{\mbox{\boldmath $\gamma$}}^{\dagger}}_{#1}}

\begin{document}
\title{{Kitaev ring threaded by a magnetic flux: Topological gap, Anderson localization of quasiparticles, and divergence of supercurrent derivative}}

\author{Martina Minutillo}
\affiliation{Dipartimento di Fisica ``E. Pancini'', Università di Napoli Federico II, Complesso di Monte S. Angelo, via Cinthia, I-80126 Napoli, Italy}

\author{Procolo Lucignano}
%\href{mailto:gpiccitto@sissa.it}{gpiccitto@sissa.it} \\
\affiliation{Dipartimento di Fisica ``E. Pancini'', Università di Napoli Federico II, Complesso di Monte S. Angelo, via Cinthia, I-80126 Napoli, Italy}

\author{Gabriele Campagnano}
\affiliation{CNR-SPIN, c/o Complesso di Monte S. Angelo, via Cinthia, I-80126 Napoli, Italy}

\author{Angelo Russomanno}
\affiliation{Scuola Superiore Meridionale, Università di Napoli Federico II, Largo San Marcellino 10, I-80138 Napoli, Italy}
\affiliation{Dipartimento di Fisica ``E. Pancini'', Università di Napoli Federico II, Complesso di Monte S. Angelo, via Cinthia, I-80126 Napoli, Italy}

\begin{abstract}
	We study a superconducting Kitaev ring  pierced by a magnetic flux, with and without  disorder, in a quantum ring configuration, and {in} a rf-SQUID one, where a weak link is present. In the rf-SQUID configuration, in the topological phase, the supercurrent shows jumps at specific values  of the flux $\Phi^*=\frac{hc}{e}(1/4+n)$, with $n\in\mathbb{N}$. In the thermodynamic limit $\Phi^*$ is constant inside the topological phase, independently of disorder, and we analytically predict this fact using a perturbative {approach} in the weak-link {coupling}. The weak link breaks the topological ground-state degeneracy, and opens a spectral gap for $\Phi\neq \Phi^*$, that vanishes at $\Phi^*$ with a cusp providing the current jump. Looking at the quasiparticle excitations, we see that they are Anderson localized, so they cannot carry a resistive contribution to the current, {and the localization length shows a peculiar behavior at a flat-band point for the quasiparticles}. In the absence of disorder, we analytically and numerically find that the chemical-potential derivative of the supercurrent logarithmically diverges at the topological-to-trivial transition, in agreement with the transition being {of the} second order.
\end{abstract}
\maketitle
\section{Introduction}
Topological quantum phase transitions have been one of the main {research areas} in the last decades~\cite{royal,Rod,Bern}. Quantum phase transitions  are characterized by  a local order parameter that becomes nonvanishing and infinite-range correlated (long-range order)~\cite{Sachdev} in the ordered phase. Topological  phase  transitions in contrast are characterized  by  a global rearranging, that is witnessed by a nonlocal string order parameter: In the thermodynamic limit the expectation of the infinite string becomes nonvanishing (see examples in~\cite{chen,ueda,berg,esposti,berg1,fazio}). 

Particularly interesting are the properties of the topological superconductors~\cite{Alicea_2012,Bern}. What is remarkable in these systems is {the closure} of the gap at the boundary between the topological superconductor and the trivial vacuum, and the appearance there of zero-energy Majorana modes~\cite{Alicea_2012,Leijnse_2012}. 
They were  first predicted, in the case of a spinless one dimensional $p$-wave superconductor~\cite{Kitaev_2001}, and are very important, both from an intrinsic theoretical point of view -- as they realize an old prediction by Majorana~\cite{major} -- and due to the possible applications in the field of quantum information~\cite{Leijnse_2012}.

Because of their importance many methods have been put forward in order to observe them. Majorana modes should exhibit distinctive experimental signatures in the tunneling conductance~\cite{stan,seng,prada,prada1,law,flens}, ballistic point contact conductance~\cite{wim}, Coulomb blockade spectroscopy~\cite{L_Fu, zazu,hut,vanh,chiu,cao}, and Josephson current~\cite{fu_kane,lut,oreg,domio, sau, peng, vayr, virta, houze, badi, san, pikulin, domi, jiang, law, van, 1, ios,procolo:prb2012,procolo:prb2013,PhysRevB.94.205125}.  

In particular, considering a clean Kitaev wire in the form of a ring pierced by a magnetic flux $\Phi$, and putting in it a weak link, one should see a characteristic jump discontinuity in the supercurrent at a value of the flux that in the thermodynamic limit tends to $\Phi^*=\Phi_0/4$, where $\Phi_0 = 2\pi\hbar c/e$ is the flux quantum~\cite{Nava_2017}. Also other models of topological superconductors show, in the same setting, discontinuities of the current when the flux is varied, but the position in flux of these discontinuities depends on the parameters of the model, also in the thermodynamic limit~\cite{Marra_2016}.

\begin{figure}
    \includegraphics[width=80mm]{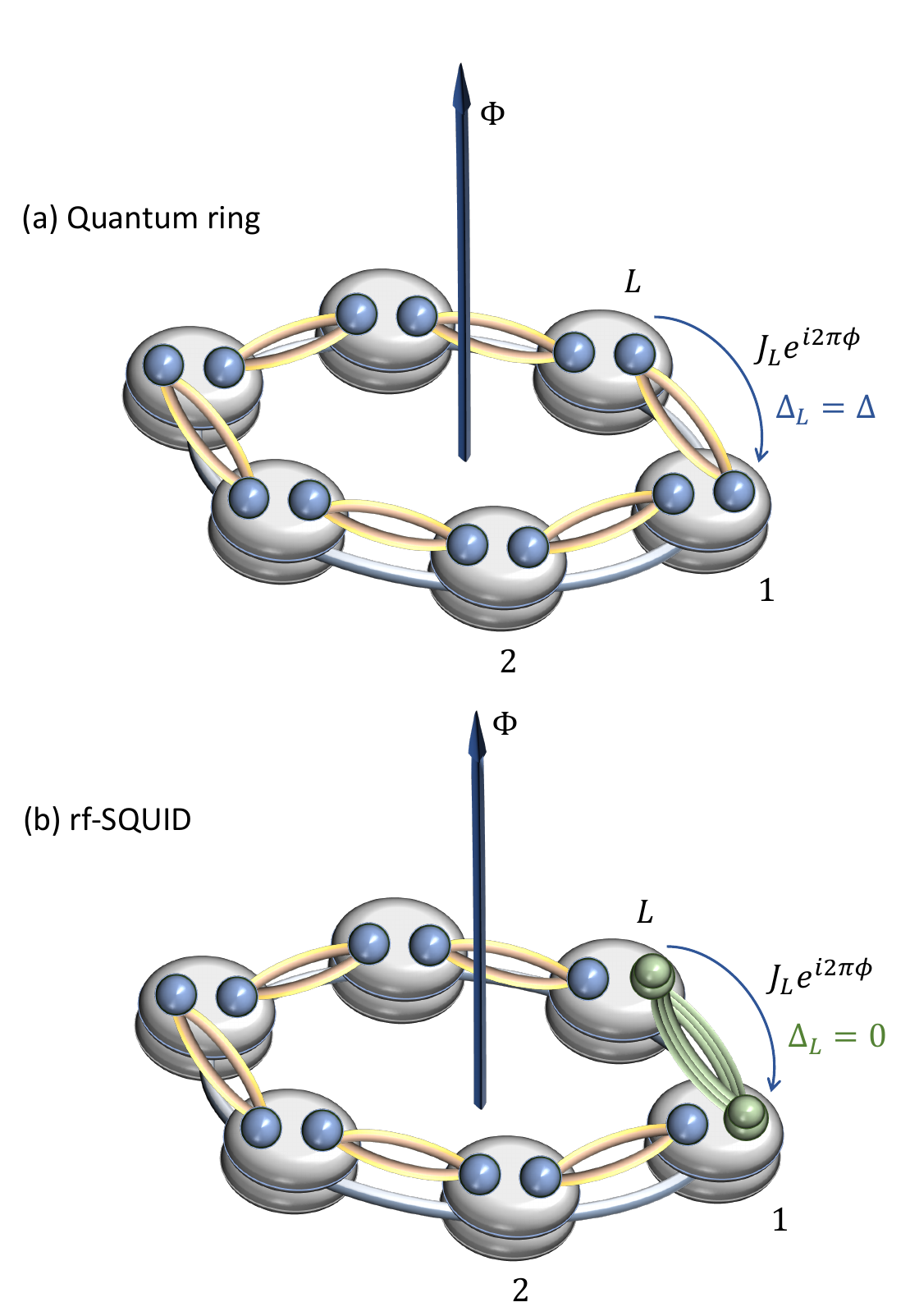}
    \caption{Schematic representation of the 1D Kitaev ring pierced by magnetic flux in the topological phase, in the quantum ring (a) and rf-SQUID (b) configuration. In each lattice site represented by the grey ovals, the Majorana fermions (blue spheres) are bound in pairs located on the neighboring sites, also those at the ends of the chain, due to the ring configuration (described by the hopping term $J_{L} e^{i 2 \pi \phi}$). The bonds are represented by the yellow ties. In (b), in the topological phase, the weak link couples the Majoranas represented by the green doubled spheres, leading to the splitting discussed in Sec.~\ref{jj:sec}. }
    %{[Per favore, metti $J_L$ ai termini di boundary]}}
    \label{fig_model}
\end{figure}

%In this manuscript we propose an indirect method of detection of  Majorana fermions, in the sense that we can detect the position of the topological-to-trivial transition. In order to do that, 
We consider a Kitaev model in a ring geometry pierced by a magnetic flux.~\cite{Nava_2017,procolo:prb2013}. We focus on the ground state in two different configurations. 
First, we we consider an uninterrupted ring with periodic boundary conditions in the superconducting terms [quantum ring -- Fig.~\ref{fig_model}(a)], second include a weak link in the ring, mimicking an rf-SQUID [Fig.~\ref{fig_model}(b)]. In both configurations we numerically find a persistent current in the ground state, that depends on $\Phi$ and tends to a finite limit for large system size. This is physically a supercurrent, indeed if we remove the superconducting terms this current vanishes in the large-size limit. We study the properties of the supercurrent, both for the case of clean system, and in the presence of disorder.

{We find that the jump of the current in the flux in the rf-SQUID configuration (observed before in [48]) is a direct consequence of the existence of Majorana excitations that take a finite energy due to the weak link. With a simple perturbative approach we are able to predict that the weak link opens a gap between the two topologically degenerate ground states (the one with and the one without Majorana modes). This topological gap closes at the topological phase transition and clearly marks this transition also in the presence of disorder. On the opposite, in the usual open-chain setting, spectral-gap features do not in general allow to distinguish the transition from the topological to the trivial phase in the disordered case (at least at finite size -- see Sec.~\ref{jj:sec}). The topological gap closes up for specific values of the external field [$\Phi^*=\Phi_0(1 / 4 + n)$ with $n\in\mathbb{N}$] with a cusp, that gives rise to a jump in the current. In the thermodynamic limit $\Phi^*$  does not directly depend on the presence and form of disorder, and is only a direct effect of the Majorana modes.}
    
     {We find that also in the presence of disorder there is a condensate that can carry a supercurrent if the flux is not vanishing. On the opposite, in presence of disorder quasiparticle excitations are Anderson localized, and so there is no normal contribution to the current, if the temperature becomes nonvanishing.  At $\Delta = J$, the localization length of the quasiparticles remains finite even for very small disorder. The zero-disorder limit is singular due to the flat-band feature of the disorderless unperturbed quasiparticles, and this behavior is very different from the normal case without superconductivity. On the opposite, outside this interval, the localization length of the quasiparticles behaves more or less as in the normal case, and diverges as a power law in the limit of vanishing disorder.}
    
    {In the clean case we numerically and analytically see that the derivative of the current with respect to the chemical potential logarithmically diverges at the transition point. This is in agreement with the fact that the topological-to-trivial transition in the Kitaev model is second order. This finding is interesting because so far there were no physical quantities in the Kitaev model diverging at the transition and so witnessing it. Some of these quantities were known in the spin representation of the Kitaev model (the quantum Ising chain in transverse field). A well-known example is the magnetic susceptibility, but has no direct physical meaning for the Kitaev model, in contrast with the current derivative we consider here.}

{We emphasize that the logarithmic divergence of the current derivative at the critical point is a robust property, and appears in both configurations.
Some} other properties of the supercurrent depend on the chosen configuration. For instance in the quantum ring the current is periodic in $\Phi$ of period $\Phi_0$ while in the rf-SQUID the periodicity is $\Phi_0/2$ in the large-size limit. In both cases the result is physically meaningful: The current aims to screen the magnetic flux, so that the resulting flux in the ring becomes quantized in units of the elementary flux $\Phi_0/2$ (see for instance~\cite{PhysRevLett.7.43,PhysRevLett.7.51,tinkham}). %Our model is too rough, and we do not consider the magnetic flux induced by the current, so we do not see this screening effect.

The paper is organized as follows. In Sec.~\ref{treat:sec} we introduce the Kitaev Hamiltonian, and the methods to treat it numerically and analytically also for large system sizes, thanks to the fact that it is a fermion quadratic model. In Sec.~\ref{dirty:sec} we discuss the model in presence of disorder. We provide results on the of the topological spectral gap in the rf-SQUID in Sec.~\ref{jj:sec}, results on the corresponding jumps in the supercurrent in Sec.~\ref{cucud:sec}, and results on the localization of the quasiparticle excitations in Sec.~\ref{loco:sec}. In Sec.~\ref{clean:sec} we show our numerical results for the clean model, and discuss how the logarithmic divergence at the critical point of the derivative in the chemical potential of the current can be predicted analytically, in the quantum ring configuration, for $2|\sin(\pi\Phi/\Phi_0)|\ll 1$.  In Sec.~\ref{conc:sec} we draw our conclusions.
%..................................................................................................................................................................................................%
\section{The Hamiltonian and its numerical treatment}\label{treat:sec}
%..................................................................................................................................................................................................%
\subsection{Definition of the Hamiltonian and unitary transformation}\label{defi:sec}
We consider the following Kitaev Hamiltonian with magnetic flux
\begin{align} \label{hammer:eqn}
    \hat{H} &= -\sum_{j=1}^{L-1}\left(J\nep^{i\varphi_j}\opcdag{j}\opc{j+1}+{\rm H.~c.}\right)-J_L\left(\nep^{i\varphi_L}\opcdag{L}\opc{1}+{\rm H.~c.}\right)\nonumber\\
    &-\sum_{j=1}^{L-1}\left(\Delta\opcdag{j}\opcdag{j+1}+{\rm H.~c.}\right){-\Delta_L\left(\hat{c}_{L}^\dagger\hat{c}_{1}^\dagger+{\rm H.~c.}\right)}+\sum_{j=1}^L\mu_j\opcdag{j}\opc{j}\,,    
\end{align}
with periodic boundary conditions $\opc{L+1}=\opc{1}$. If not otherwise specified, we take in the boundary term $J_L=J$. The onsite chemical potential is $\mu_j$; We define $h_j=\mu_j/2$ in order to get simpler formulae. {We have also  introduced the parameter $\Delta_L$ at the boundary. It can acquire the value $\Delta_L=0$ in the rf-SQUID configuration [Fig.~\ref{fig_model}(a)], or the value $\Delta_L=\Delta$ in the quantum ring one [Fig.~\ref{fig_model}(b)]. The $\varphi_j$ are the Peierls phases that we discuss later.}

{The $h_j$ are local fields. In Sec.~\ref{dirty:sec} we take random $h_j$ uniformly distributed in the interval $[-h,h]$. In the absence of magnetic flux, the system is topological when~\cite{Kitaev_2001,fisher_prb95}
\begin{equation}\label{cond:eqn}
    {\log J} >  \overline{\log |h_j|}\,,
\end{equation}
where $\overline{(\ldots)}$ marks the average over disorder realizations. The topological phase corresponds to the magnetized phase of the quantum Ising chain in transverse field~\cite{fisher_prb95}, to which this model is mapped by means of the Jordan-Wigner transformation~\cite{lieb,PFEUTY197079}. In the disordered case, using numerical methods similar to the ones used in~\cite{strinati_prb17} for the cluster-Ising model, one can show that there is string order if condition Eq.~\eqref{cond:eqn} is obeyed.}

{In Secs.~\ref{clean:sec},~\ref{analy:sec} we take the parameters uniform ($h_j\equiv h$) so that the model without flux shows a second-order transition at $h=J$~\cite{Kitaev_2001,PFEUTY197079}.
%Using the Jordan-Wigner mapping~\cite{lieb} to the quantum Ising chain in transverse field~\cite{PFEUTY197079}, one can see that the phase for $h<J$ (topological phase) is characterized by a nonvanishing expectation of a nonlocal string order parameter (that in the spin representation is a simple spin correlator)~\cite{string_order_note}. 
Thanks to the mapping to the quantum Ising chain in transverse field, the critical properties of this transition are well known~\cite{Sachdev,scaletta}; In particular this transition is second order and one can see discontinuities in the second derivatives of the ground-state energy (for instance the magnetic susceptibility in the Ising representation).}

 {The hopping terms contain the Peierls phase
\begin{equation}\label{phas1:eqn}
    \varphi_j = \frac{e}{\hbar c}\int_{j}^{j+1}{\bf A}\cdot\ud l\,.
\end{equation}}
%
%Because the superconducting terms are such that $\Delta_j = \braket{\opcdag{j}\opcdag{j+1}}$, the phase of $\Delta$ and that of the operators $\opcdag{j}\opcdag{j+1}$ cancel each other, so we can take $\Delta$ with no phase and uniform. % ({Questa cosa non la capisco bene. Perch\'e $\Delta$ non dipende da $j$ nel caso con disordine?}) Any phase change on the $\opc{j}$ will therefore keep the superconducting term unchanged, because any phase increase of the $\opc{j}\opc{j+1}$ will be exactly compensated by a phase decrease in $\Delta_j$.
%
%\begin{equation}\label{phas2:eqn}
%    \theta_j+\theta_{j+1} = -\frac{e}{\hbar c}\int_{1}^{j}{\bf A}\cdot\ud l-\frac{e}{\hbar c}\int_{1}^{j+1}{\bf A}\cdot\ud l\,.
%\end{equation}

We choose a magnetic flux $\Phi$ piercing the ring formed by the system (see Fig.~\ref{fig_model}). %Let us imagine the system as a ring with circumference $L$ and radius $L/(2\pi)$. The vector potential will be given by
%
%\begin{equation} \label{Aphi:eqn}
%    {\bf A}=\frac{\Phi}{2\pi r}\hat{\boldsymbol{\theta}} =
%    \frac{\Phi}L\hat{\boldsymbol{\theta}}\,,
%\end{equation}
%
%where $\hat{\boldsymbol{\theta}}$ is the tangential versor. Substituting this formula into 
Applying Eq.~\eqref{phas1:eqn}  we find
\begin{equation}
    \varphi_j = \varphi = 2\pi\frac{\Phi}{\Phi_0 L} %nonumber\\
%    \theta_j+\theta_{j+1} &= -2\pi\frac{\Phi}{\Phi_0 L}(2j-1)\,,
\end{equation}
where we have defined the {flux quantum $\Phi_0=\frac{2\pi\hbar c}{e}$} and emphasized the independence on $j$.

%The magnetic flux is encoded in the phases $\theta_j$. 
%
%while the superconducting terms contain the superconducting phase. % ($\Delta$ is indeed imposed from the outside in this kind of system) %. 

%\subsection{Time-independent case}\label{unitary_ti:sec}
%
In order to locally eliminate the phases in Eq.~\eqref{hammer:eqn} we apply the following unitary transformation
\begin{equation}\label{trasfo1:eqn}
    \widetilde{c}_j = \opc{j}\nep^{-i\theta_j}\,,
\end{equation}
where $\theta_j = -(j-1)\varphi$. {Let us first focus on the terms $\Delta \hat{c}_j^\dagger\hat{c}_{j+1}^\dagger$ in Eq.~\eqref{hammer:eqn}, where $\Delta\equiv \braket{\hat{c}_j\hat{c}_{j+1}}$. Applying this transformation to the term $\braket{\hat{c}_j\hat{c}_{j+1}} \hat{c}_j^\dagger\hat{c}_{j+1}^\dagger$ one can easily see that this term stays unchanged.} Focusing on the hopping terms, one can see that all the phases in the Hamiltonian Eq.~\eqref{hammer:eqn} disappear, but the one of the boundary hopping term, %. Indeed, noticing that %
%\begin{equation} \label{thetus:eqn}
%    \theta_j = -(j-1)\varphi\,,
%\end{equation}
%
and the resulting Hamiltonian has the form
\begin{align}\label{hbt:eqn}
  &\widetilde{H} = -\sum_{j=1}^{L-1}J\left(\widetilde{c}_{j}^\dagger\widetilde{c}_{j+1}+{\rm H.~c.}\right)-
    \sum_{j=1}^{L-1}\Delta\left(\widetilde{c}_{j}^\dagger\widetilde{c}_{j+1}^\dagger+{\rm H.~c.}\right)\nonumber\\
     &\hspace{-0.3cm}+\sum_{j=1}^L2h_j\widetilde{c}_{j}^\dagger\widetilde{c}_{j}-J_L\left(\nep^{2\pi i\phi}\,\widetilde{c}_L^\dagger\widetilde{c}_1+{\rm H.~c.}\right)-\Delta_L\left(\widetilde{c}_{L}^\dagger\widetilde{c}_{1}^\dagger+{\rm H.~c.}\right)\,.
\end{align}
where the flux appears only in the phase of the boundary term where we have defined
\begin{equation}
  \phi=\frac{L\varphi}{2\pi} = \frac{\Phi}{\Phi_0}\,.
\end{equation}
% From now on, we will make reference only to the Hamiltonian Eq.~\eqref{hbt:eqn}, where we will redefine the $\widetilde{c}_j$ operators as $\opc{j}$.
%

We graphically represent this Hamiltonian in Fig.~\ref{fig_model}. On each site the two spheres mark the two Majorana modes associated to that site~\cite{Alicea_2012}. In the quantum ring configuration [Fig.~\ref{fig_model}(a)] the situation is perfectly translation invariant, while in the rf-SQUID one [Fig.~\ref{fig_model}(b)] there is the weak link, that in the topological phase couples the unpaired Majorana modes at the boundary, leading to the spectral gap described in Sec.~\ref{jj:sec}. Let us move now to describe the techniques to diagonalize this fermionic quadratic Hamiltonian.
%.........................................................................................................................%
\subsection{Diagonalization of the Hamiltonian}\label{diago:sec}
The rotated Hamiltonian Eq.~\eqref{hbt:eqn} can be written in a matrix notation as
\begin{eqnarray} \label{quadratic-H:eqn}
\widetilde{H} = \opbfPsidag{} \, {\mathbb H} \, \opbfPsi{}
%=\sum_{i,j=1}^{2L} \Psi^{\dagger}_i H_{i,j} \Psi_j
%  = \left( \begin{array}{cc}  \opbfcdag{}\!\! &,\; \opbfc{} \!\! \end{array} \right)
= \left( \, \opbfcdag{} \,,\, \opbfc{} \!\! \right)
  \left( \begin{array}{cc} {\Q} & {\B} \\
                           -{\B}^* & -{\Q}^*\end{array} \right)
     \left( \begin{array}{l}  \opbfc{} \!\! \\ \opbfcdag{} \!\! \end{array} \right) \;.
\end{eqnarray}
\begin{equation}
    {\Q} = \left(\begin{array}{cccccc}
           {h}_1&-\frac{J}2&0&0&\ldots&-\frac{J_L}2\nep^{i2\pi\phi}\\
           -\frac{J}2&{h}_2&-\frac{J}2&0&\ldots&0\\
           0&-\frac{J}2&{h}_3&-\frac{J}2&\ldots&0\\
           \vdots&\vdots&\vdots&\vdots&\vdots&\vdots\\
           0&      0    & 0    & \ldots&{h}&-\frac{J}2\\
           -\frac{J_L}2\nep^{-i2\pi\phi}&0&0&\ldots&-\frac{J}2&{h}_L\,,
           \end{array}\right)
\end{equation}
%
%where we define
%
%\begin{equation}
%    \widetilde{h}_j=h_j + \pi\frac{v}{\Phi_0}\frac{j-1}{L}\quad{\rm for}\quad j = 1,\,\ldots,\,L\,,
%\end{equation}
%
and
\begin{equation}
    {\B} = \left(\begin{array}{cccccc}
           0&-\Delta/2&0&0&\ldots&-\Delta_L/2\\
           \Delta/2&0&-\Delta/2&0&\ldots&0\\
           0&\Delta/2&0&-\Delta/2&\ldots&0\\
           \vdots&\vdots&\vdots&\vdots&\vdots&\vdots\\
           0&      0    & 0    & \ldots&0&-\Delta/2\\
           \Delta_L/2&0&0&\ldots&\Delta/2&0
           \end{array}\right)\,.
\end{equation}
Following~\cite{glen}, we see that the ground state is defined by the eigenvalue equation
\begin{equation}\label{dyndondan:eqn}
    \left( \begin{array}{cc} {\Q} & {\B} \\
                           -{\B}^* & -{\Q}^* \end{array} \right)\left( \begin{array}{cc} {\U} & {\V}^* \\
                           {\V} & {\U}^* \end{array} \right)=\mathbb{E}_d \left( \begin{array}{cc} {\U} & {\V}^* \\
                           {\V} & {\U}^* \end{array} \right)\,,
\end{equation}
where $\mathbb{E}_d$ is the diagonal matrix with eigenvalues $(\epsilon_1,\,\ldots,\,\epsilon_L,\,-\epsilon_1,\,\ldots,\,-\epsilon_L)$ (these eigenvalues are called Bogoliubov spectrum). At the cost of reorganizing the matrices, we can take all the $\epsilon_\mu\geq 0$. The unitary transformation
\begin{equation} \label{trasfo:eqn}
    \left( \begin{array}{l}  \opbfc{} \!\! \\ \opbfcdag{} \!\! \end{array} \right) = \left( \begin{array}{cc} {\U} & {\V}^* \\
                           {\V} & {\U}^* \end{array} \right) \left( \begin{array}{l}  \opbfgamma{} \!\! \\ \opbfgammadag{} \!\! \end{array} \right)
\end{equation}
applies, and ground state $\ket{\psi}$ is defined as the one annihilated by all the fermionic $\opgamma{}$ operators
\begin{equation}\label{annihilator:eqn}
    \opgamma{\mu}\ket{\psi}=0\quad\forall\;\mu=1,\,\ldots,\,L\,.
\end{equation}
The ground-state energy is provided by
\begin{equation}\label{eggs:eqn}
  \mathcal{E} = -\sum_{\mu=1}^L\epsilon_\mu\,.
\end{equation}
The $2\epsilon_\mu$ have furthermore the interpretation as fermionic quasiparticle energy excitations (with corresponding creation operator $\opgammadag{\mu}$). Indeed, the Hamiltonian can be written as
\begin{equation}
  \widetilde{H} = \sum_{\mu=1}^L\epsilon_\mu(\opgammadag{\mu}\opgamma{\mu}-\opgamma{\mu}\opgammadag{\mu})\,,
\end{equation}
and it is easy to see that, whatever the choice of $\mu$, $\opgammadag{\mu}\ket{\psi}$ is still an eigenstate with energy $\mathcal{E}'=\mathcal{E}+2\epsilon_\mu$. We can physically interpret it as the condensate with a fermionic quasiparticle excitation on top. Notice that we can write the quasiparticle creation operator as
\begin{equation}\label{amp:eqn}
  \opgammadag{\mu} = \sum_j U_{j\,\mu}^* \opcdag{j} + V_{j\,\mu}\opc{j}\,,
\end{equation}
so the $U_{j\,\mu}^*$ and $V_{j\,\mu}$ acquire the physical meaning of probability amplitude of the quasiparticle, a property that will be useful for inquiring localization properties.
%..................................................................................................................................................%
\subsection{Current operator}
The current operator is defined as~\cite{thouless,supp}
\begin{equation}
    \label{current_operator}
    \hat{I} = \frac{c}{\Phi_0}\dfrac{\partial \hat{H}}{\partial \phi} = - 2\pi i\frac{c}{\Phi_0}J_L \left[ e{^{2\pi i \phi}}\opcdag{L}\opc{1} - e^{-2\pi i {\phi}}\opcdag{1}\opc{L} \right]  \, .
\end{equation}
\noindent Then, by evaluating the current expectation on the ground state defined as in (Eq.~\eqref{annihilator:eqn}), we can write the expectation of the current as
%
%\begin{equation}
%    \label{expectation_current}
%    \braket{\psi|I|\psi} = -i \frac{2\pi c}{\Phi_0}\frac{J}{L}\sum_{j=1}^{L}\sum_{\mu=1}^{L} \left[e{^{2\pi i \frac{\phi}{L}}} v^*_{j,\mu} v_{j+1, \mu} - e{^{-2\pi i \frac{\phi}{L}}} v^*_{j+1,\mu} v_{j, \mu} \right]   \, .
%\end{equation}
%
%In the rest of the notes we will take $c=1$ and $2\pi/\Phi_0 = e/(\hbar c ) = 1$. Performing the same analysis in the new representation we get  - e{^{-2\pi i {\phi}}} v^*_{1,\mu} v_{L, \mu}
%
\begin{equation}
    \label{expectation_current1}
   I=\braket{\psi|I|\psi} = -i 2\pi\frac{cJ_L}{\Phi_0}\sum_{\mu=1}^{L} 2\Im{e{^{2\pi i {\phi}}} v^*_{L,\mu} v_{1, \mu} }  \, ,
\end{equation}
and this formula is valid for both choices of boundary conditions. We emphasize that, evaluating the expectation over $\ket{\psi}$ of the current operator Eq.~\eqref{current_operator} and applying the Hellmann-Feyman theorem (see for instance~\cite{Grosso}), one can write the expectation of the current as
\begin{equation}\label{ig:eqn}
I=\frac{c}{\Phi_0}\dfrac{\partial \mathcal{E}(\phi)}{\partial \phi}=-\frac{c}{\Phi_0}\sum_{\mu=1}^L\dfrac{\partial \epsilon_\mu(\phi)}{\partial \phi}\,,
\end{equation}
where we have explicited the flux dependence in the ground-state energy Eq.~\eqref{eggs:eqn}. 

From now on in the numerics we will express everything in units of $J$ ($J=1$). {Moreover we will measure velocity in units of $c$ and flux in units of $\Phi_0/(2\pi)$.}
%We can consider two different types of boundary conditions in the superconducting terms. We can consider open boundary conditions, where there is no
%.....................................................................................................................................................................................................%
\section{Disordered (dirty) case} \label{dirty:sec}
Let us now consider in Eq.~\eqref{hbt:eqn} the case of a disordered chemical potential, with $h_j$ uniformly distributed in the range $ - h \leq h_{j} \leq h$. Using Eq.~\eqref{cond:eqn} we see that the transition occurs for 
\begin{equation}\label{tradis:eqn}
  h = \operatorname{e} J\,,
\end{equation}
 where $\operatorname{e}$ is the Neper number.

We will numerically perform averages over $N_{s}$ realizations of the disorder, {indicating} them with the symbol $\overline{(\ldots)}$. We choose $N_s$ {large} enough that the errorbars are negligible~\cite{nota_errorbar}.

We consider the topological properties of the spectral gap in Sec.~\ref{jj:sec}, and Anderson localization of quasiparticles in Sec.~\ref{cucud:sec}. Throughout all this section, where not otherwise specified, we take $\Delta = J = 1$.
%-----------------------------------------------------------------------------------------------------------------------------------------------------------------------------------------------------%
\subsection{Topological gap in the rf-SQUID} \label{jj:sec}
Let us inquire the effect of the Majorana modes in the rf-SQUID configuration, where in the topological phase these modes exist at the boundary corresponding to the weak link. %, and have an effect on the physics. (In the quantum ring there is no boundary and no Majorana modes).
%There is another important difference between the two choices of boundary conditions, that we can see in the topological phase ($h<1$). From one side, the OBC case shows jumps at $\phi=1/4 + n/2$, where $n$ is integer (with a deviation exponentially small in $L$~\cite{Nava_2017}), as we can see in Fig.~\ref{fig:varie}(d). From the other, the current in the PBC case is perfectly continuous [Fig.~\ref{fig:varie}(a)]. The point is that this discontinuity is an effect due to the presence of Majorana modes, that can exist only in OBC, because there is a boundary to accommodate them, due to the presence of the weak link. T

\begin{figure}
    \centering
    \begin{tabular}{c}
    (a)\\
     \hspace{0cm}\includegraphics[width=80mm]{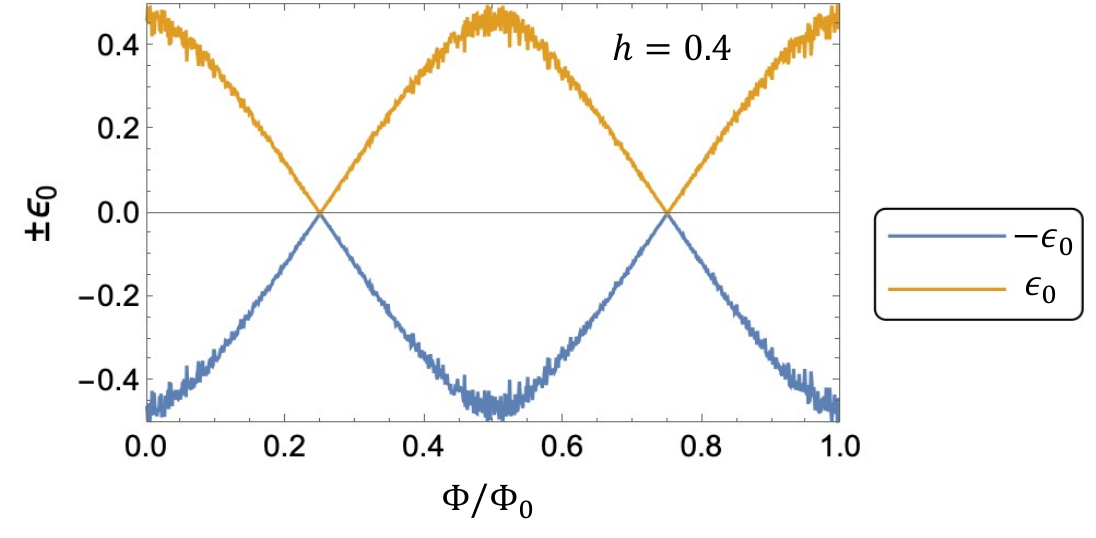}\\
    (b)\\
     \hspace{0cm}\includegraphics[width=80mm]{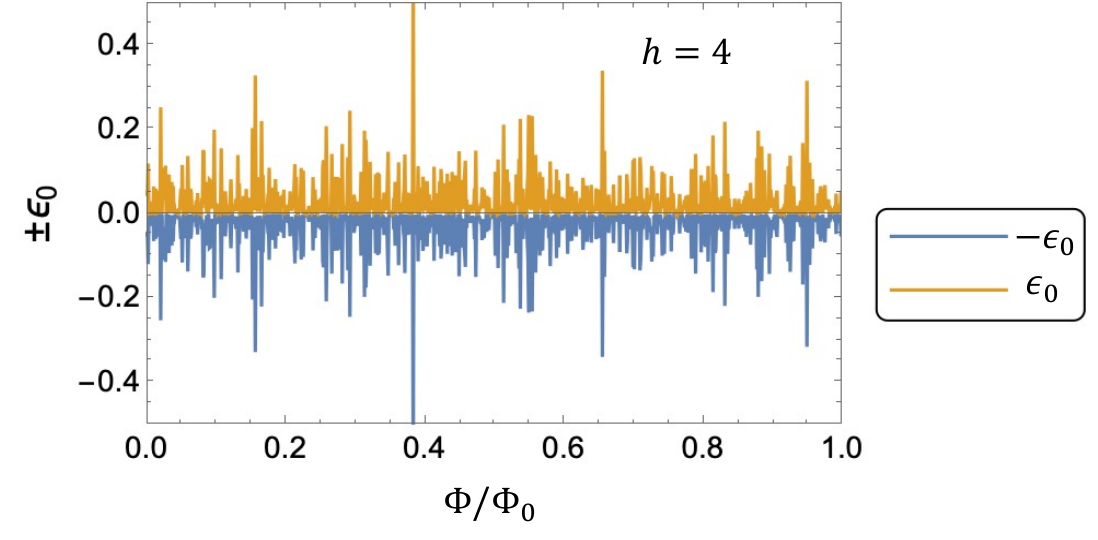}
    \end{tabular}  
    \caption{{Lowest Bogoliubov energy $\epsilon_0$ and its opposite versus $\phi$ for a single disorder realization in the rf-SQUID configuration, for parameters in the topological phase (panel a), and the trivial phase (panel b). {Recall that the gap between the ground state and the first excited one is $\Delta\epsilon = 2\epsilon_0$.} Notice in panel (a) the degeneracies at $\phi=1/4+n/2$ and the corresponding cusps in the lower branch that give rise to the jumps in the supercurrent. Numerical parameters: $L=50$, $h=0.4$ for panel (a), $L=50$, $h=4$ for panel (b).}}
    \label{fig:cusp_dirty}
\end{figure}

{Majorana modes have physical effects first of all on the spectrum. In case of open chain the Majorana modes lead to a ground state that is  doubly degenerate in the thermodynamic limit. The weak link changes the situation and breaks this degeneracy, opening a spectral gap everywhere but at $\phi=1/4 + n/2$, as we can see in Fig.~\ref{fig:cusp_dirty}(a). Here we show the two central energies $\pm \epsilon_0$ of the Bogoliubov spectrum versus $\phi$. We can see that a gap opens between them whenever $\phi\neq 1/4+n/2$. We remark that this gap is a topological effect and disappears outside the topological phase [see Fig.~\ref{fig:cusp_dirty}(b)].}

 {This gap (that we call topological) corresponds to the lowest quasiparticle energy $2\epsilon_0$ being nonvanishing. So there is a gap also in the many-body spectrum, and the ground-state double degeneracy is broken. The double degeneracy is restored only for $\phi=1/4+n/2$, where the lowest-energy quasiparticle (actually a Majorana mode) has vanishing energy in the thermodynamic limit, and acting on the ground state creates a state degenerate with it.}
 
  This peculiar behavior is a direct effect of the Majorana modes, as a perturbative approach, valid in the thermodynamic limit, shows.
%
%\begin{figure}[!b]
%    \centering
%    \includegraphics[width=80mm]{{imgs_6_06/evals_vs_phi_L_50_mu_0.4_OBC}.pdf}
%    \caption{ Quasiparticle energies $\pm\epsilon_0$ corresponding to the Majorana modes. Notice the degeneracies at $\phi=1/4+n/2$ and the corresponding cusps in the lower branch that give rise to the jumps in the supercurrent. Numerical parameters: $L=50$, $h=0.4$.}
%    \label{cusp:fig}
%\end{figure}

%\begin{figure}[h!]
%    \centering
%    \begin{tabular}{c}
%     \hspace{0cm}\includegraphics[width=80mm]{imgs_6_06/evals_vs_phi_L_50_mu_0.4_OBC_dirty}
%    \end{tabular}  
%    \caption{Quasiparticle energies $\pm\epsilon_0$ corresponding to the Majorana modes, for a single disorder configuration. Notice the degeneracies at $\phi=1/4+n/2$ and the corresponding cusps in the lower branch that give rise to the jumps in the supercurrent. Numerical parameters: $L=50$, $h=0.4$. {[Sull'asse delle ascisse metti $\pm\epsilon_0$ e anche nella legenda. Per confronto metti anche un pannello con un caso triviale.]}}
%    \label{fig:cusp_dirty}
%\end{figure}

% In the thermodynamic limit the position of the jump is provided by a simple argument. 
The argument goes as follows. Let us first consider the open chain ($J_L=\Delta_L=0$)%without flux
\begin{eqnarray}\label{h0:eqn}
  &\hat{H}_0 =  -\sum_{j=1}^{L-1}J\left(\hat{c}_{j}^\dagger\hat{c}_{j+1}+{\rm H.~c.}\right)\nonumber\\
   &-\sum_{j=1}^{L-1}\Delta\left(\hat{c}_{j}^\dagger\hat{c}_{j+1}^\dagger+{\rm H.~c.}\right)+\sum_{j=1}^L2h_j\hat{c}_{j}^\dagger\hat{c}_{j}\,,
\end{eqnarray}
and call $\ket{\rm GS}$ its ground state. In the topological phase [Eq.~\eqref{cond:eqn}] there are two Majorana modes at the boundaries of the chain, the left one $\gamma_\ell$ and the right one $\gamma_R$, such that in the thermodynamic limit the states $\ket{\psi_1}=\frac{1}{2}(\gamma_\ell+i\gamma_R)\ket{\rm GS}$ is degenerate with the ground state~\cite{Alicea_2012}. 

The weak-link term $\hat{V} = -J_L\left(\nep^{2\pi i\phi}\,\hat{c}_L^\dagger\hat{c}_1+{\rm H.~c.}\right)$ does not couple these two states, but has a different expectation on each of them. In particular we can write
{
\begin{align}
  \braket{{\rm GS}|\hat{V}|{\rm GS}}&=2A\cos(2\pi\phi)\,,\nonumber\\
  \braket{\psi_1|\hat{V}|\psi_1}&= 2B\cos(2\pi\phi)\,,
\end{align}
with $A=-J_L\braket{{\rm GS}|\hat{c}_L^\dagger\hat{c}_1|{\rm GS}}$ and  $B=-J_L\braket{\psi_1|\hat{c}_L^\dagger\hat{c}_1|\psi_1}$ both real, being the Hamiltonian in Eq.~\eqref{h0:eqn} invariant under time reversal.} [The reality of $B$ can be seen most clearly for $\mu=0$, where $\frac{1}{2}(\gamma_\ell+i\gamma_R) = \frac{1}{2}(\opc{1}+\opcdag{1}+\opc{L}-\opcdag{L})$~\cite{Alicea_2012}]. 
% $\braket{{\rm GS}|\hat{V}|{\rm GS}}=-J\Re{\braket{{\rm GS}|\hat{c}_L^\dagger\hat{c}_1|{\rm GS}}\nep^{2\pi i\phi}}$. Being the Hamiltonian in Eq.~\eqref{h0:eqn} invariant under time reversal, $\braket{{\rm GS}|\hat{c}_L^\dagger\hat{c}_1|{\rm GS}}$ is real, and so  $\braket{{\rm GS}|\hat{V}|{\rm GS}}=A\cos(2\pi\phi)$, with $A=-J_L\braket{{\rm GS}|\hat{c}_L^\dagger\hat{c}_1|{\rm GS}}$ real. For the same reason $\braket{\psi_1|\hat{V}|\psi_1}= B\cos(2\pi\phi)$, with $B=-J_L\braket{\psi_1|\hat{c}_L^\dagger\hat{c}_1|\psi_1}$ real. [The reality of $B$ can be seen most clearly for $\mu=0$, where $\frac{1}{2}(\gamma_\ell+i\gamma_R) = \frac{1}{2}(\opc{1}+\opcdag{1}+\opc{L}-\opcdag{L})$~\cite{Alicea_2012}]. 

So, if $J_L\ll 1$ and then the perturbative theory is valid, a gap opens between the two degenerate ground states and is given by  %the smallest excitation over the ground state -- that in the topological phase is a Majorana excitation --  has the approximate formula valid for $J_L\ll 1$
\begin{equation}\label{qp:gap}
  \Delta \epsilon=|B-A||\cos(2\pi\phi)|\,.
\end{equation}
In particular the gap vanishes for $\phi=1/4 + n/2$, where the states  $\ket{\rm GS}$ and $\ket{\psi_1}$ become degenerate. In order to compare with the numerics shown in Fig.~\ref{fig:cusp_dirty}, let us recall that $\Delta \epsilon = 2\epsilon_0$, that's to say the gap is equal to the smallest quasiparticle excitation energy, as we have discussed above. Comparing with Fig.~\ref{fig:cusp_dirty}(a) we see that Eq.~\eqref{qp:gap} correctly predicts the points where $\epsilon_0$ vanishes, the presence of cusps at these points, and the periodicity in $\phi$, also outside its regime of validity (in Fig.~\ref{fig:cusp_dirty} $J_L=J=1$).%Up to terms exponentially small in $L$ (that in the thermodynamic limit are vanishing) we have $\braket{\psi_\ell|\hat{V}|\psi_\ell}=\braket{\psi_R|\hat{V}|\psi_R}=0$, and $\braket{\psi_\ell|\hat{V}|\psi_R}=-J\Re{\braket{{\rm GS}|\gamma_\ell\hat{c}_L^\dagger\hat{c}_1\gamma_R|{\rm GS}}\nep^{2\pi i\phi}}$. Because the Hamiltonian Eq.~\eqref{h0:eqn} is invariant under time inversion, the expectation $A\equiv \braket{{\rm GS}|\gamma_\ell\hat{c}_L^\dagger\hat{c}_1\gamma_R|{\rm GS}$ is real. This can be most clearly seen in the limit $h\to 0$, where $\gamma_\ell (that can be $\ket{\rm GS}$ or $\ket{\psi_1}$, according to $\phi$)

{So the interplay of Majorana modes with the flux gives rise to a topological spectral gap that does not exist in the trivial phase.}
The presence of the topological gap for $\phi\neq 1/4 + n/2$ is an important information, because gives a way to probe the topological phase in the rf-SQUID setting. In order to do that, we fix $\phi = 1/2$ and plot the disorder-averaged gap $\overline{\Delta\epsilon} \equiv 2\overline{\epsilon_0}$ versus $h$ [Fig.~\ref{fig:Log(Devals)_vs_h_ALLOPEN}(a)]. We see that there is a minimum that, when the system size increases, becomes sharper and smaller, and its position moves towards the transition point {$h=\nep$ [see Eq.~\eqref{tradis:eqn}.}

To emphasize the relevance of this result, we plot the same averaged gap for the case of open chain ($J_L=\Delta_L=0$)  [Fig.~\ref{fig:Log(Devals)_vs_h_ALLOPEN}(b)]. Here the gap is much smaller in the topological phase than in the trivial one, but there is no feature that sharply marks the transition. {[The situation is different for the clean case -- see Appendix~\ref{app:gap}.]}
%
%So, a finite jump in the current is strictly associated with the topological phase, while the jump identically disappears in the trivial phase. In the limit of large $L$ we find that the jump continuously vanishes at the transition, as we can see that in Fig.~\ref{jump_vs_h:fig} (curve for $L=800$), where we plot the height of the jump $\Delta I$ versus $h$. For smaller size $L$, in contrast, there is a discontinuity in $\Delta I$ near the transition. For $L=800$ we have checked that for $h$ approaching $J$ from below, the jump linearly vanishes as $\Delta I \sim |J-h|$, so that the critical exponent is 1. This gives rise to a singularity in the derivative $\partial \Delta I /\partial h$ that is discontinuous at the transition.
%

%.....................................................................................................................................................................................................%
\subsection{Current jumps in the rf-SQUID}\label{cucud:sec}
The dependence of $\epsilon_0$ on $\phi$ in the topological phase of the rf-SQUID, as shown in Fig.~\ref{fig:cusp_dirty}, allows to make a prediction on the properties of the ground-state current. The current is provided by Eq.~\eqref{ig:eqn}, so is proportional to the sum of the $\partial \epsilon_\mu(\phi) / \partial\phi$. But one of the $\epsilon_\mu$ (the $\epsilon_0$)  has a cusp for $\phi=1/4 + n/2$ [see Fig.~\ref{fig:cusp_dirty} and Eq.~\eqref{qp:gap}]. 

Applying the derivative in $\phi$ we get therefore a discontinuity of the ground-state current for $\phi=1/4 + n/2$ in the topological phase, and this is precisely what we observe [see Fig.~\ref{I_phi_dirty:fig}(c)]. These jumps disappear in the quantum ring configuration [Fig.~\ref{I_phi_dirty:fig}(a,b)], and outside the topological phase [Fig.~\ref{I_phi_dirty:fig}(b,d)], where there are no Majorana modes and no cusps in $\epsilon_0$.
%
%\\
\begin{widetext}
\begin{figure*}
    \centering
    \begin{tabular}{cc}
    (a)&(b)\\
     \hspace{0cm}\includegraphics[width=82mm]{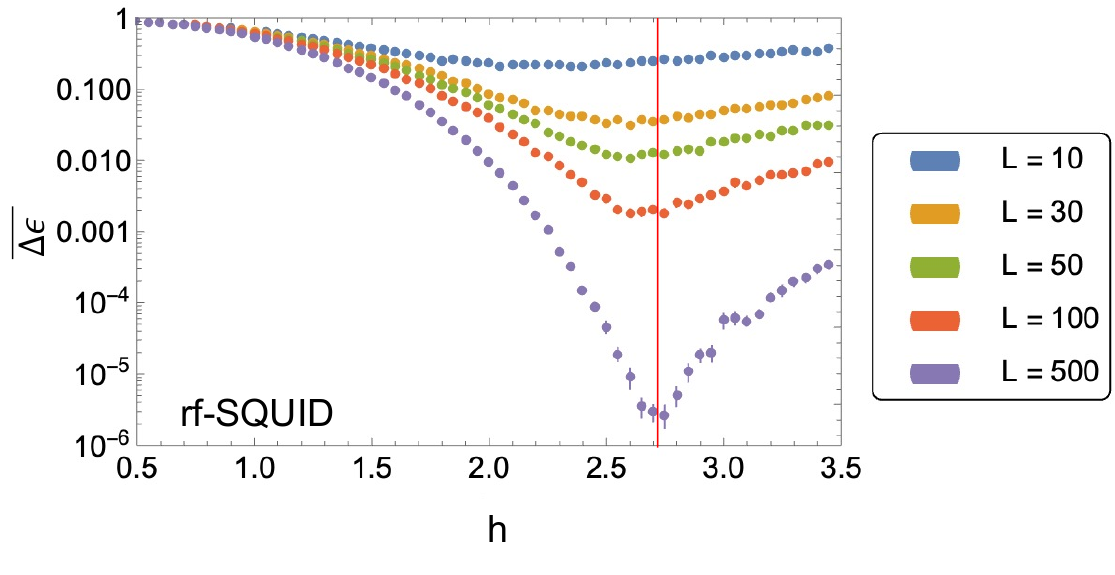}&
     \hspace{0cm}\includegraphics[width=82mm]{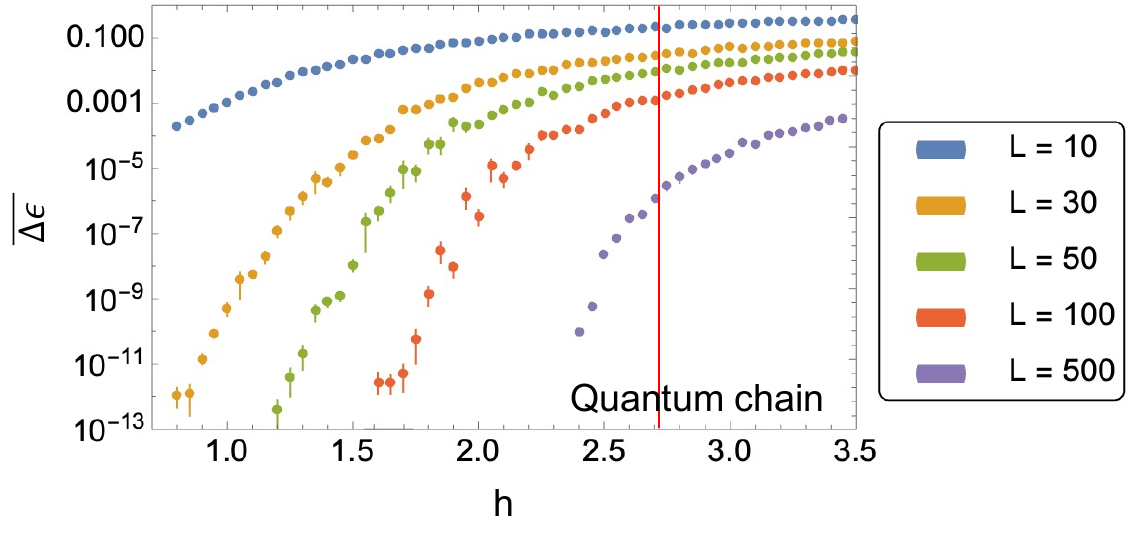}
    \end{tabular}  
    \caption{Disorder-averaged gap $\overline{\Delta \epsilon}$ versus $h$ for the rf-SQUID configuration (panel a), and for a open quantum superconducting chain (panel b),  for different $L$. {The vertical line at $h=\nep$ marks the topological-to-trivial transition point. In panel (a) notice the minimum marking the transition.} Numerical parameters: $N_{s}=1000$, $\phi=0.5$ for panel (a).} 
    %{[Aggiungi le barre verticali al punto teorico di transizione $h=\nep$.]}}  and $\phi=0.4$ for panel (b)
    \label{fig:Log(Devals)_vs_h_ALLOPEN}
\end{figure*}
\begin{figure*}
    \centering
    \begin{tabular}{cc}
    (a) & (b) \\
     \hspace{0cm}\includegraphics[width=80mm]{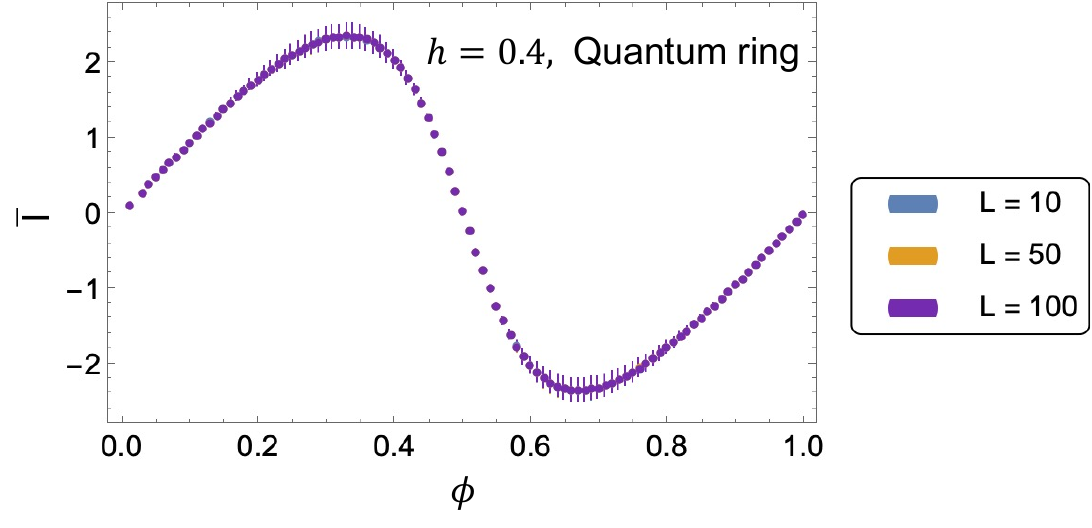}&\includegraphics[width=80mm]{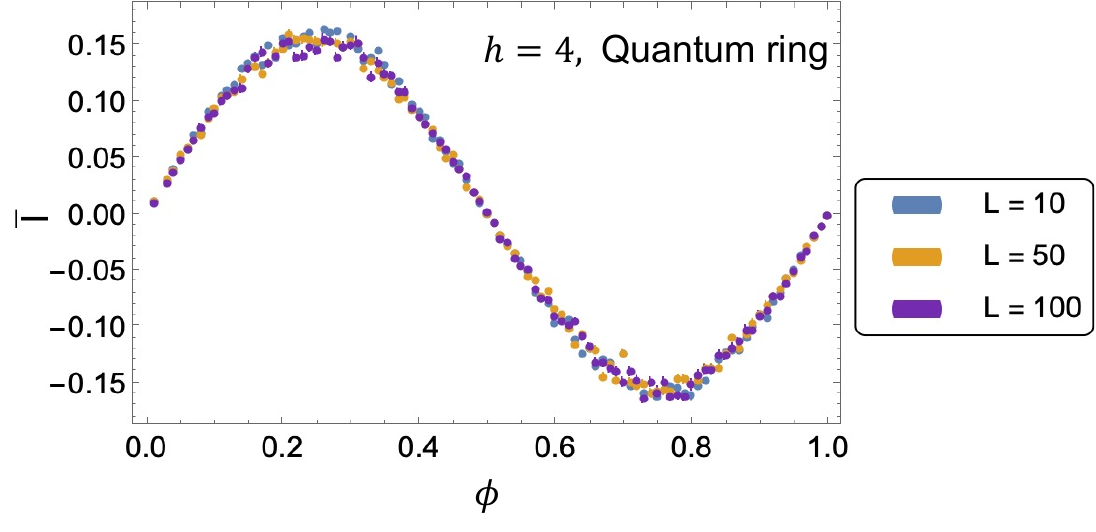}\\
     (c) & (d) \\
     \includegraphics[width=80mm]{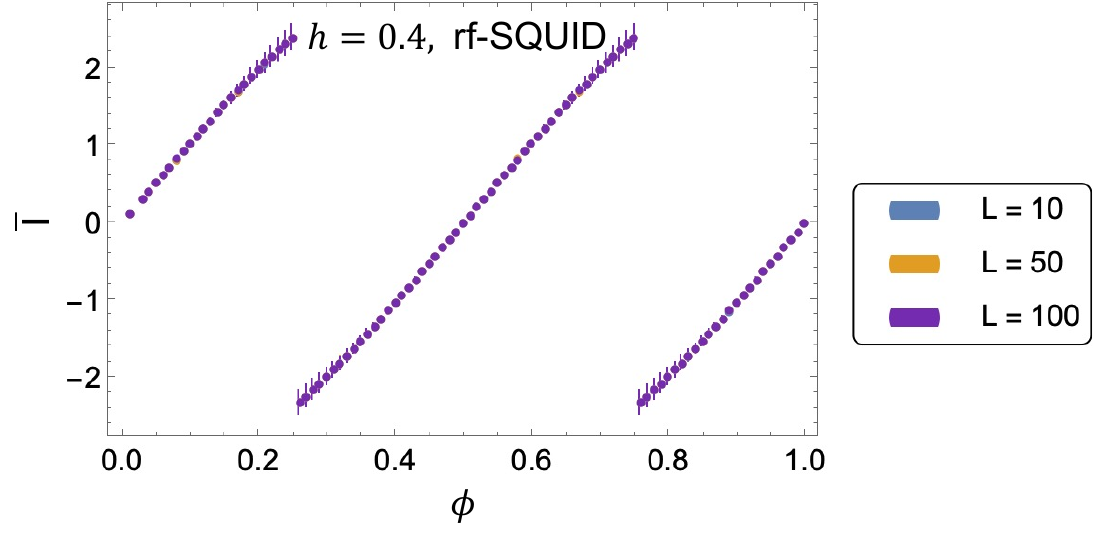}&\includegraphics[width=80mm]{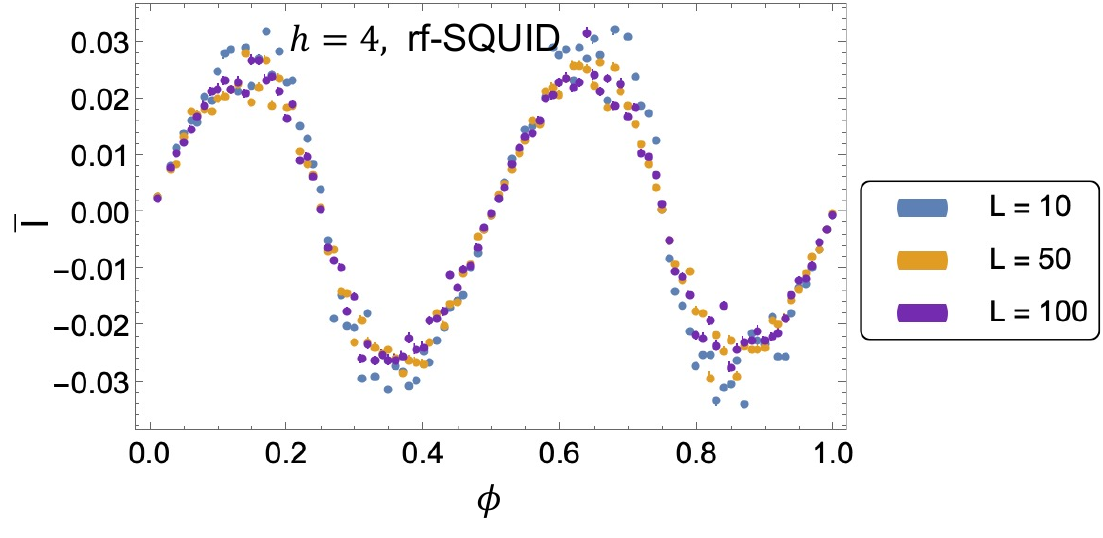}
    \end{tabular}
         
    \caption{(Left) Current versus $\phi$ for different system size $L$. On the top row (panels a, b) we take the quantum ring configuration, on the bottom row (panels c, d) the rf-SQUID. On the left column (panels a, c) we take $h=0.4$, on the right column (panels b, d) we take $h=4$. Notice the current jumps at $\phi=\frac{1}{4}+\frac{n}{2}$ in panel c. $N_{s} = 1000$.} 
    %{(Cambia PBC con quantum ring e OBC con rf-SQUID. Metti $h=4$ al posto di $h=2.7$. Metti le barre d'errore. Metti $\phi$ sull'asse orizzontale.)}
    \label{I_phi_dirty:fig}
\end{figure*}
\begin{figure*}
    \centering
    \begin{tabular}{cc}
    (a)&(b) \\
     \hspace{-0.5cm}\includegraphics[width=80mm]{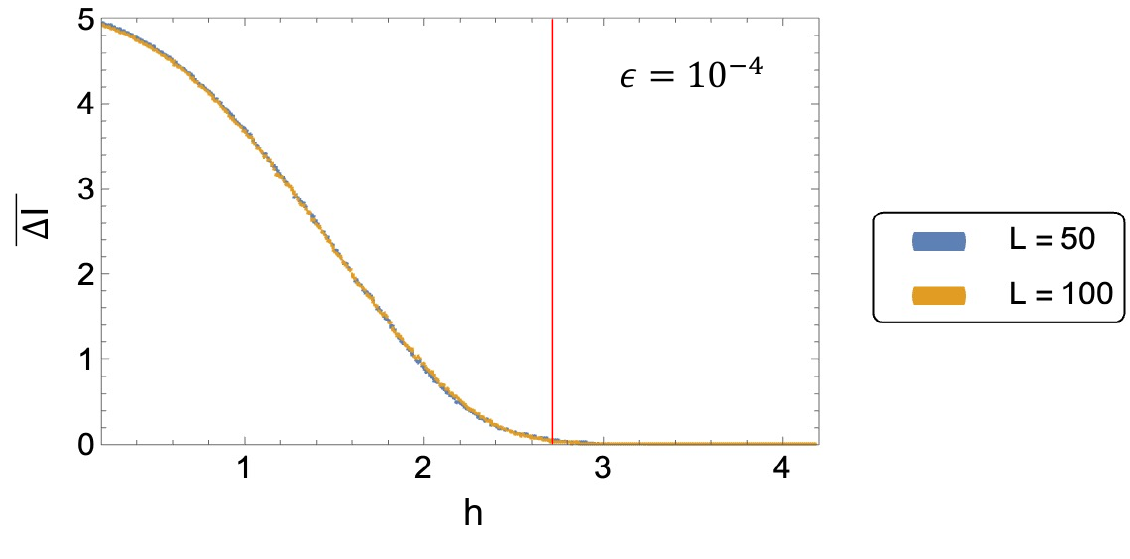}&
     \hspace{0.5cm}\includegraphics[width=85mm]{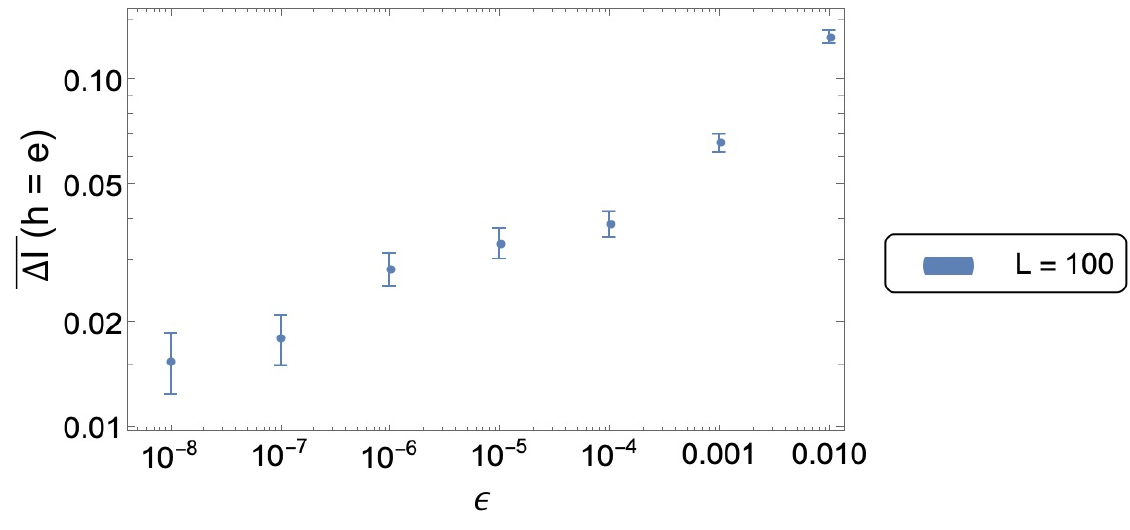}
    \end{tabular}
    \caption{(Panel a) Jump in the current $\Delta\overline{I}$ [Eq.~\eqref{OK:eqn}] versus $h$ in the rf-SQUID, for different system size $L$.  The vertical red line indicates the theoretical transition point at $h = \operatorname{e} $.{ (Panel b) $\Delta\overline{I}$ versus $\epsilon$ at the transition point (double logarithmic plot).} Numerical parameters: $N_{s}=2000$,  $\epsilon=10^{-4}$ in panel (a).}
    %{Vedi com'è se si mette la scala logaritmica sull'asse verticale.}
    \label{fig:DI_vs_h_OBC_DIRTY}
\end{figure*}
\end{widetext}

The jumps in the current in the rf-SQUID configuration were already observed in~\cite{Nava_2017}, and their position was analytically predicted for the clean case. Here, numerically and analytically, we show that  in the thermodynamic limit their position is always $\phi=1/4 + n/2$, independently of the form of the disorder. Thanks to this robustness, we can use the current jump at $\phi=1/4 + n/2$ to probe the topological phase. We do that in Fig.~\ref{fig:DI_vs_h_OBC_DIRTY}(a), where we show the averaged current jump versus $h$. {We numerically evaluate it as
\begin{equation}\label{OK:eqn}
  \Delta\overline{I} = \overline{I}\left(\phi=\frac{1}{4}+\epsilon\right) -   \overline{I}\left(\phi=\frac{1}{4}-\epsilon\right)\,.
\end{equation}
This is just an approximation and tends to the jump in the limit $\epsilon\to 0$. In Fig.~\ref{fig:DI_vs_h_OBC_DIRTY}(a) we have taken $\epsilon=10^{-4}$ and we can see that $\Delta\overline{I}$ is slightly nonvanishing just above the transition $h=\nep$. The point is that convergence in $\epsilon$ is slower near the transition. Nevertheless, taking $\epsilon$ small enough, $\Delta\overline{I}$ converges to 0 also at the transition point, approximately as a power law, as we show in Fig.~\ref{fig:DI_vs_h_OBC_DIRTY}(b).}
%
%see that it is nonvanishing inside the topological phase, and becomes vanishing slightly above the theoretical transition point $h=\nep$.
%

%To be precise, the position of the jumps is  $\phi=1/4 + n/2$ only in the thermodynamic limit, at any finite size we see deviations that become stronger as one gets near the critical point.
%

We conclude noticing that the disorder-averaged supercurrent shows a periodicity in $\Phi$ of $\Phi_0$ when the system is in the quantum ring configuration [Fig.~\ref{I_phi_dirty:fig}(a,b)], while a periodicity of $\Phi_0/2$ in the rf-SQUID one  [Fig.~\ref{I_phi_dirty:fig}(c,d) -- in the trivial phase one has to go to the large-size limit to see it fully developed]. This association between periodicity and configuration is very robust, and we always observe it (also in the clean case considered in Sec.~\ref{clean:sec}). 

So, even in the presence of disorder the superconducting system shows a persistent current in the ground state for large sizes, at variance with the normal system ($\Delta = 0$) considered in Appendix~\ref{normal:app}. The situation is different for the excitations on the ground state, that turn out to be Anderson localized, as we discuss in the next section.
%
%..............................................................................................................................................................................................%
\subsection{Localization of the quasiparticle excitations}\label{loco:sec}
In the presence of disorder, noninteracting systems show Anderson localization, a localization phenomenon of the excitations due to destructive interference induced by disorder~\cite{muller2016disorder,RevModPhys.57.287,PhysRev.109.1492}. %We are dealing with an integrable Hamiltonian, so also in our case the quasiparticle excitations defined in Eq.~\eqref{amp:eqn} might be localized in space.
 
In order to understand if {this phenomenon occurs also in our case}, we evaluate the inverse participation ratio~\cite{Edwards_1972,weg} (IPR) of the {quasiparticle-excitations} probability amplitudes [see Eq.~\eqref{amp:eqn}]. The IPR is a standard measure of localization, and in our case we start considering the IPR of a quasiparticle
\begin{equation}\label{ormo:eqn}
  {\rm IPR}_\mu = \sum_j ||U_{j\,\mu}|^2+|V_{j\,\mu}|^2|^2\,.
\end{equation}
{[The spinor $(U_{j\,\mu}^*,\,V_{j\,\mu})$ marks the space amplitude of the excitations -- see Eq.~\eqref{amp:eqn}]. }It is easy to see that if the quasiparticle is localized ($U_{j\,\mu}$, $V_{j\,\mu}$ are nonvanishing only on one or a few sites) this quantity does not scale with the system size. Instead if the quasiparticle is fully delocalized ($U_{j\,\mu}$, $V_{j\,\mu}$ are nonvanishing over an extensive number of sites with normalization $\sum_j|U_{j\,\mu}|^2+|V_{j\,\mu}|^2 = 1$) the IPR scales as $1/L$. %(There are also other scalings corresponding to a fractal behaviour of the amplitude -- see~\cite{scardicchio2017perturbation,RevModPhys.57.287}). 
The physical meaning of the IPR is an estimate of the inverse of the localization length of the quasiparticle.

\begin{figure}
    \centering
    \begin{tabular}{c}
    (a)\\
     \hspace{0cm}\includegraphics[width=80mm]{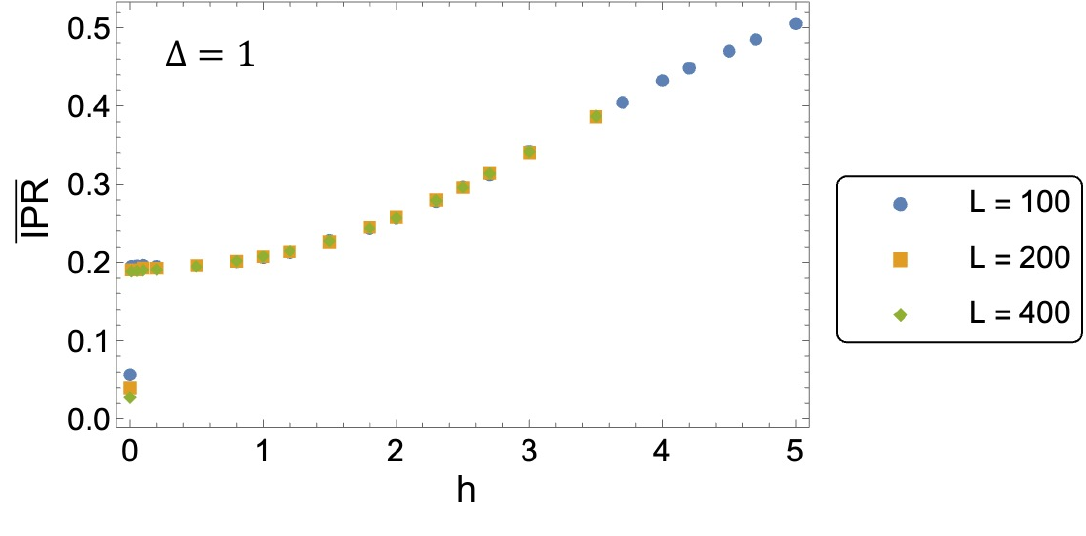}\\
     (b)\\
     \hspace{0cm}\includegraphics[width=80mm]{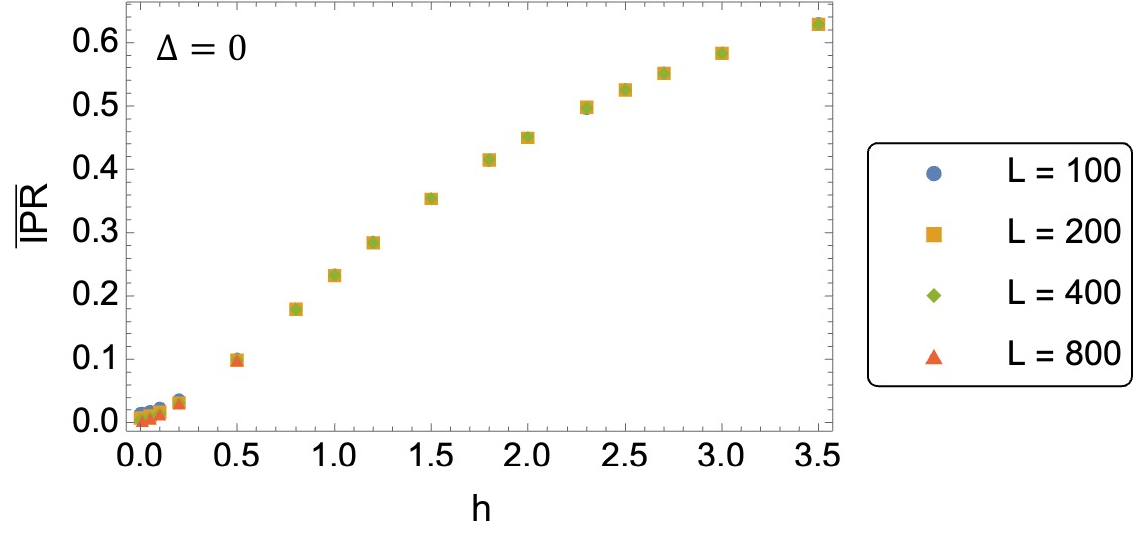}
    \end{tabular}  
    \caption{(a) $\overline{\rm IPR}$ versus $h$ in the rf-SQUID for $\Delta = 1$ and different system sizes $L$. (b) The same for $\Delta=0$. Numerical parameters: $\phi = 0.6$,  $N_{s}=500$.}
    %\textcolor{green}{Per favore, cambia la caption sull'asse delle ordinate e scrivi IPR}.
    \label{fig:IPR_vs_h}
\end{figure}

 In order to probe the scaling of the IPR, we average Eq.~\eqref{ormo:eqn} over the quasiparticles and the $N_s$  disorder realizations, and consider
\begin{equation}\label{averIPR:eq}
  \overline{\rm IPR} = \frac{1}{L}\sum_{\mu=1}^L\overline{{\rm IPR}_\mu}\,,
\end{equation}
%
%with the symbol $\overline{(\ldots)}$ meaning the average over realizations. 
We show the $\overline{\rm IPR}$ versus $h$ for $\Delta = 1$ and different values of $L$ in Fig.~\ref{fig:IPR_vs_h}(a). Except $h=0$, where there is a clear scaling with $L$, the $\overline{\rm IPR}$ saturates to a finite value already for $L=400$. So there is Anderson localization, and it seems to appear already for values of the disorder amplitude $h$ as small as $h=0.01$. When $\Delta = 1$ the limit $h\to 0$ seems therefore to be singular. For any $h\neq 0$ the inverse participation length is finite for large $L$ (meaning a finite average inverse localization length). For $h=0$ the system becomes suddenly delocalized ($\overline{\rm IPR}$ and average inverse localization length scaling to 0 with the system size).

This is different from the case without superconductivity $\Delta = 0$ [Fig.~\ref{fig:IPR_vs_h}(b)]. Here the limit $h\to 0$ is regular, and the $\overline{\rm IPR}$ (the average inverse localization length) vanishes continuously at this point. We see in Appendix~\ref{IPR:app} that this vanishing occurs as a power law with exponent $\sim 1.2$.~\cite{otano} For large $h$, instead, both in the case $\Delta = 0$ and $\Delta = 1$, the $\overline{\rm IPR}$ increases logarithmically with $L$ (see Appendix~\ref{IPR:app}), confirming the theoretical prediction of~\cite{scardicchio2017perturbation} for the inverse localization length.
\begin{figure}
   \centering
    \begin{tabular}{c}
     (a)\\
      \hspace{0cm}\includegraphics[width=86mm]{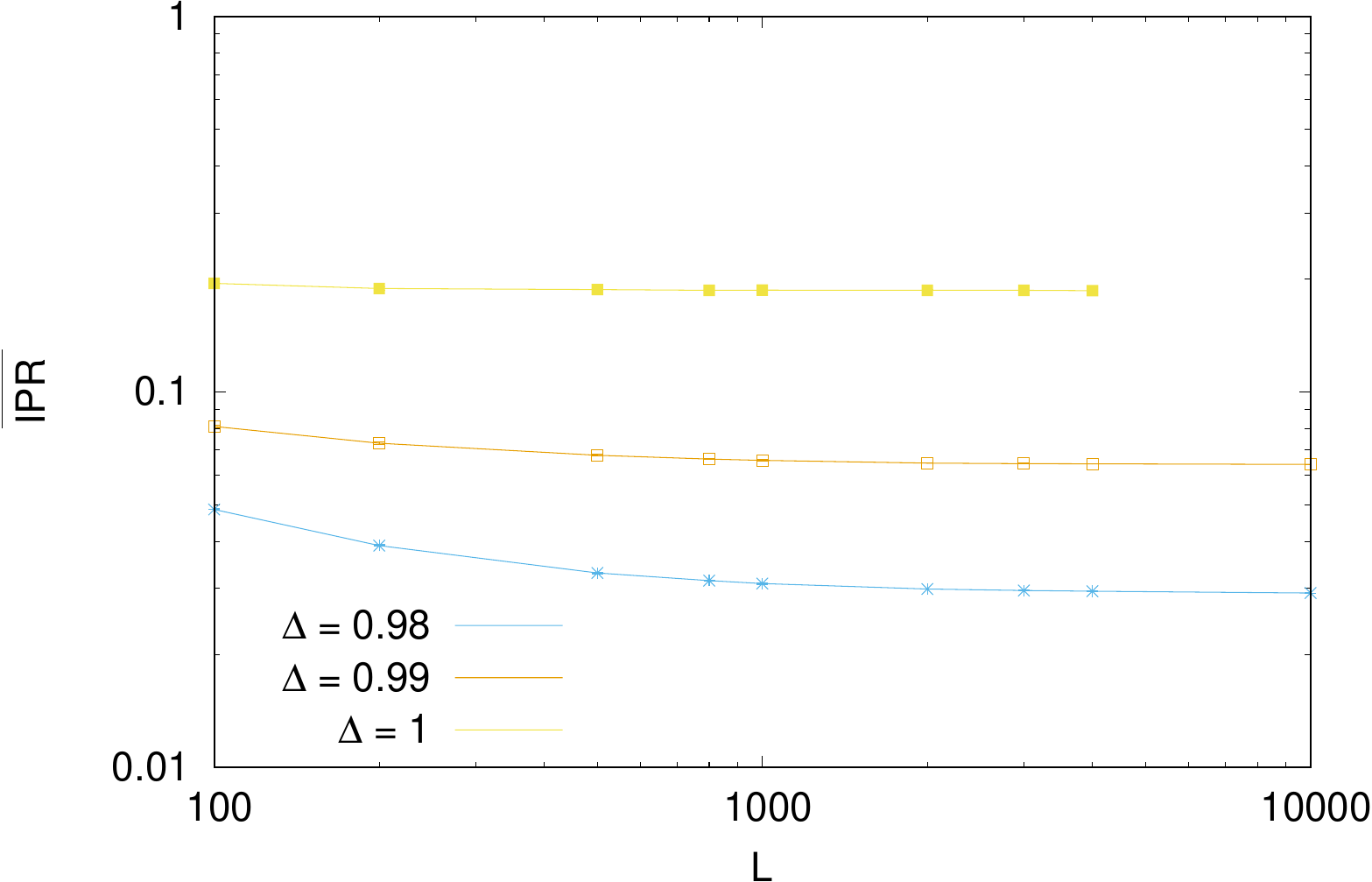}\\
    (b)\\
     \hspace{-0.35cm}\includegraphics[width=84mm]{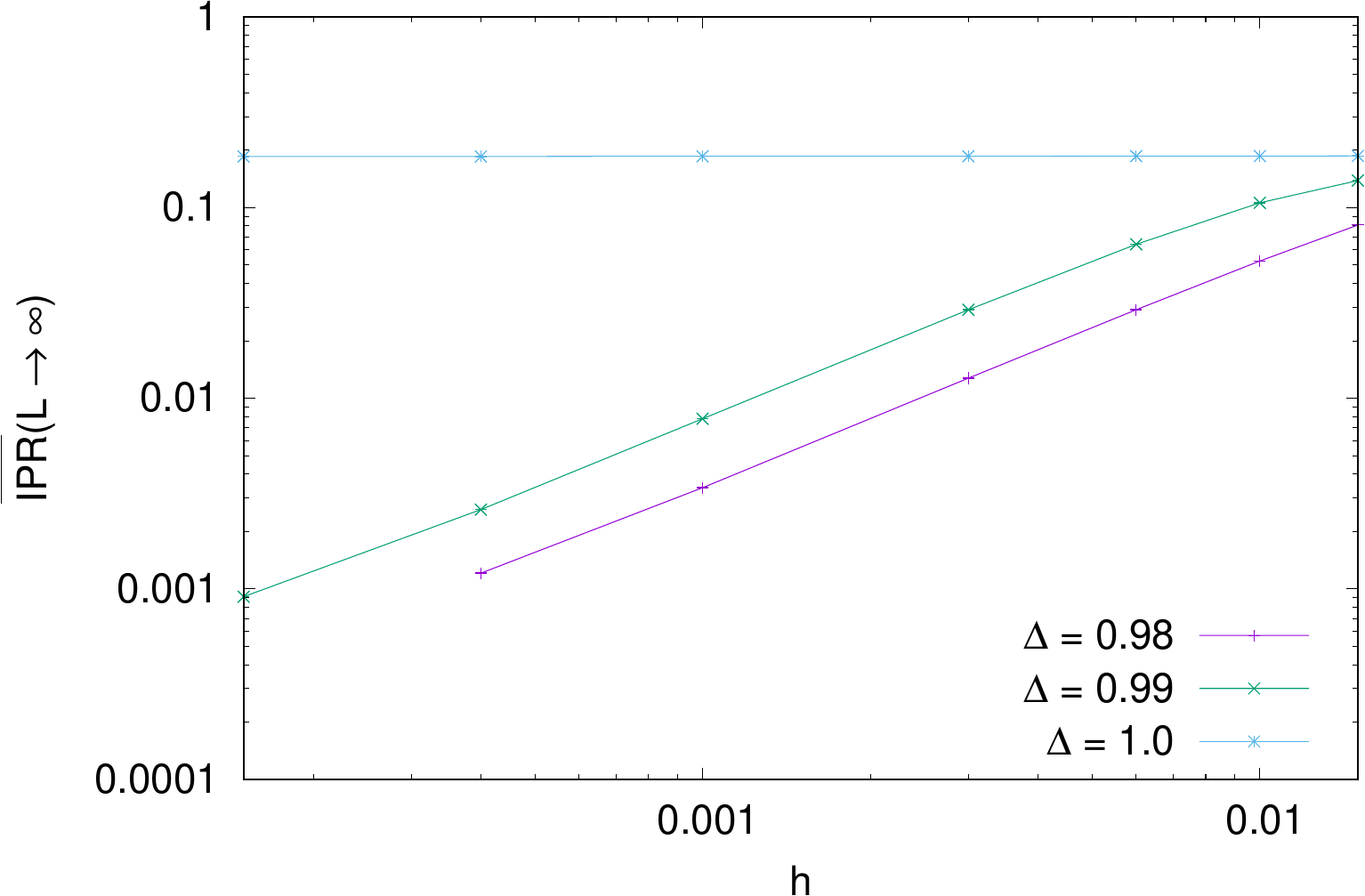}\\
    \end{tabular}  
	\caption{{(Panel a) $\overline{\rm IPR}$ versus $L$ in the quantum ring for different values of $\Delta$ and $h = 6\cdot10^{-3}$. {(Panel b) $\overline{\rm IPR}(L\to\infty)$ versus $h$ in the quantum ring different values of $\Delta$.} Notice the peculiar behavior at $\Delta = 1$. [{$\overline{\rm IPR}(L\to\infty)$ is approximated as $\overline{\rm IPR}(L=10^4)$ for $\Delta = 0.98,\,0.99$ and $\overline{\rm IPR}(L=4000)$ for $\Delta = 1$.}] Numerical parameters: $\phi = 0$, $N_{s}=48$. Errorbars are plotted but are not visible on this scale.}}
    \label{fig:Log(IPR)_vs_Delta}
\end{figure}

{In order to better understand this behavior let us consider the $\overline{\rm IPR}$ versus $L$ [see some examples in Fig.~\ref{fig:Log(IPR)_vs_Delta}(a)]. We always find that it attains a finite limit for large $L$. This finding is physically meaningful, because the system is Anderson localized, and the $\overline{\rm IPR}$ always tends to the inverse of the localization length $1/\lambda$, that is finite for any fitite $h$. Let us call the large-$L$ limit of $\overline{\rm IPR}$ as $\overline{\rm IPR}(L\to\infty)$. We plot $\overline{\rm IPR}(L\to\infty)$ versus $h$ for different values of $\Delta$ in Fig.~\ref{fig:Log(IPR)_vs_Delta}(b). For any $\Delta\neq 1$, $\overline{\rm IPR}(L\to\infty)$ scales to $0$ for $h\to 0$, similarly to what happens for the nonsuperconducting case ($\Delta = 0$). The behavior is different for $\Delta = 1$, where $\overline{\rm IPR}(L\to\infty)$ tends to a finite limit for $h\to 0$ [Fig.~\ref{fig:Log(IPR)_vs_Delta}(b)].}

{So, $\Delta = 1$ is a very special point, and is the only one where the localization length $\lambda = 1 / \overline{\rm IPR}(L\to\infty)$ stays finite in the limit of $h\to 0$. The reason of this phenomenon is the following. For $\Delta = 1,\, h= 0$ the quasiparticles are degenerate (they display a flat band with vanishing bandwidth), and even the tiniest $h\neq 0$ is enough to break this degeneracy. Because the onsite fields $h_j$ are disordered, they break the degeneracy so that the quasiparticles become localized, and the $\overline{\rm IPR}(L\to\infty)$ is finite also for the tiniest $h$. In this sense the limit $h\to 0$ is singular. For $\Delta\neq 1$ the behavior is the same as $\Delta = 1$ if $h$ is larger than the unperturbed bandwidth of the quasiparticles. When $h$ goes below the bandwidth a behavior similar to the $\Delta = 0$ case is recovered and $\overline{\rm IPR}(L\to\infty)$ scales to 0 for $h\to 0$ [see Fig~\ref{fig:Log(IPR)_vs_Delta}(b)].}
%{Looking at the behavior of the $\overline{\rm IPR}$ versus $\Delta$ for different system sizes for $h\ll J$, we see that the $\overline{\rm IPR}$ scales to 0 with $L$ everywhere but in a small region surrounding $\Delta = 1$ [see Fig.~\ref{fig:Log(IPR)_vs_Delta}(a)]. {In this region the $\overline{\rm IPR}$ tends to a finite value for increasing system size, as we show in Fig.~\ref{fig:Log(IPR)_vs_Delta}(b). Although there is the possibility that this region shrinks taking smaller $h$, the saturation behavior in $L$ we see here is in strong contrast with the decay with $L$ occurring outside this region up to at least $L=1200$ [Fig.~\ref{fig:Log(IPR)_vs_Delta}(a)].}
%
%{This peculiar behavior might be related to the fact that the point $\Delta = 1$ is very special, because for $h=0$ all the quasiparticle excitations are degenerate (in the spin representation this corresponds to a classical spin chain). However, further investigations are needed to better clarify this point.} %How this flat-band property is related to the singular limit in $h$ of the IPR is not at present to us clear, and deserves further investigation.}

%ing to a constant value for a finite value of $\Delta$ (around $\Delta \sim 0.95$ -- see Fig.~\ref{fig:Log(IPR)_vs_Delta}). So, the singular-limit situation described above develops for a finite value of $\Delta$. \\

In summary we get the interesting physical conclusion that in the disordered case the ground state carries a supercurrent (see Sec.~\ref{cucud:sec}), but the quasiparticle excitations are localized, so no normal current is possible. Moreover, for superconducting coupling $\Delta\simeq 1$, this effect occurs for the smallest disorder.
\section{Uniform (clean) case} \label{clean:sec}
\begin{widetext}
\begin{figure*}
    \centering
    \begin{tabular}{ccc}
%    \hspace{-2cm}
%    \caption{}
%    \label{fig:I_vs_phi_mu_0.4_Delta_0_L_var}
%\end{figure}
%\\
%\begin{figure}[b!]
%    \centering
    (a)&(b)&(c)\\
    \includegraphics[width=49mm]{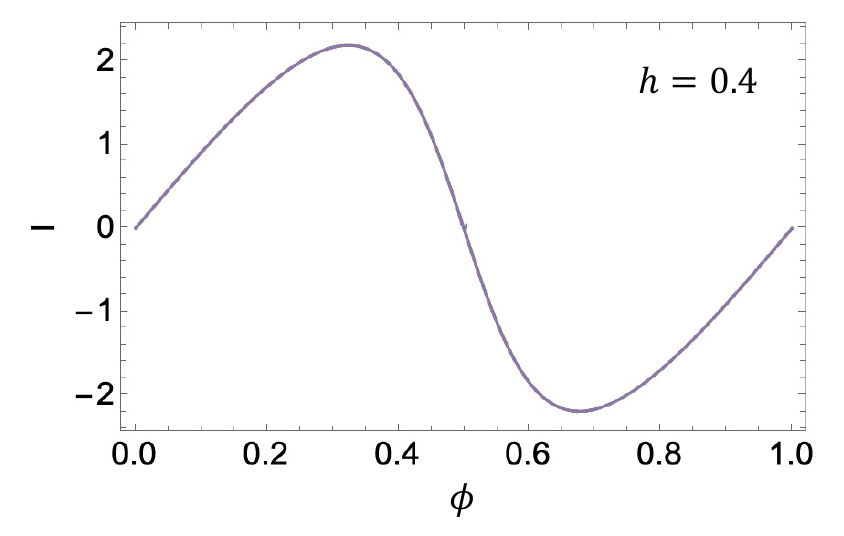}&\includegraphics[width=48mm]{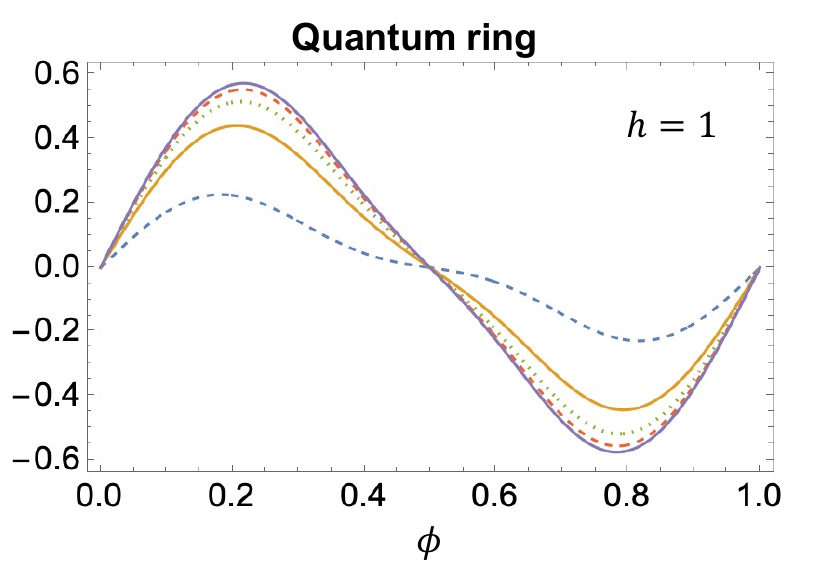}&\includegraphics[width=60mm]{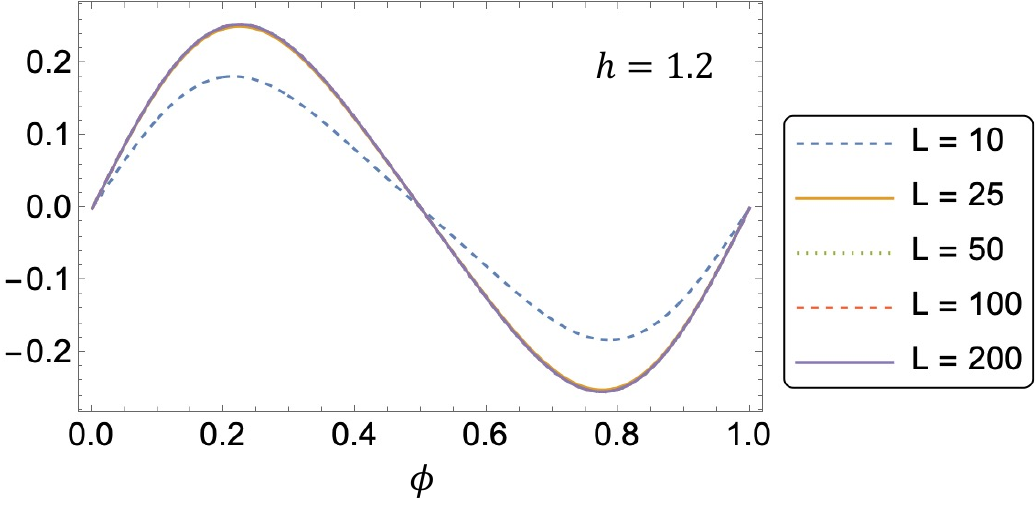}\\
    (d)&(e)&(f)\\
    \includegraphics[width=50mm]{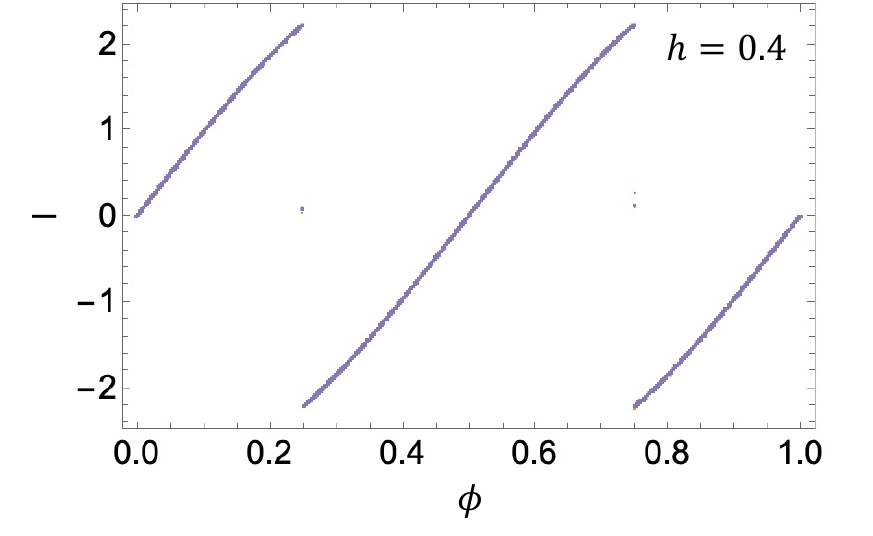}&\includegraphics[width=50mm]{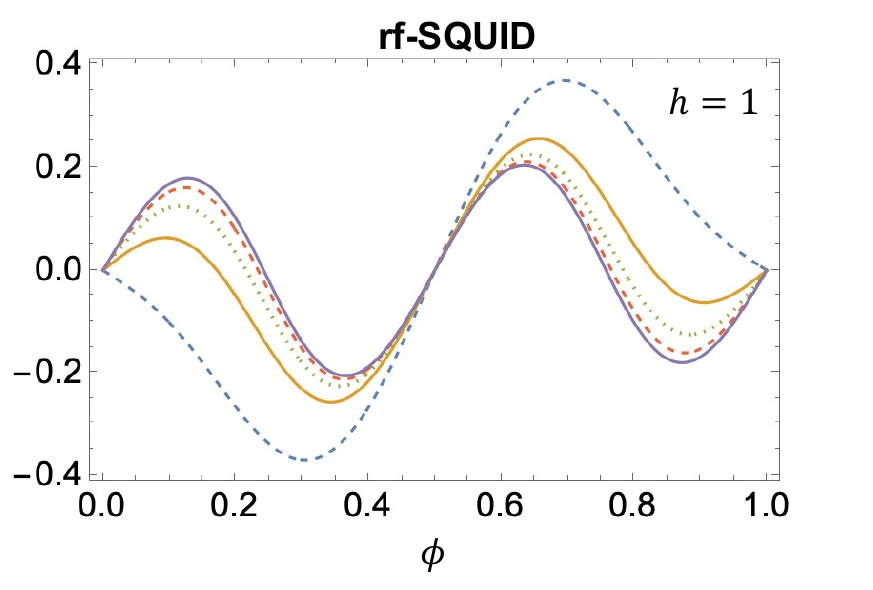}&\includegraphics[width=62mm]{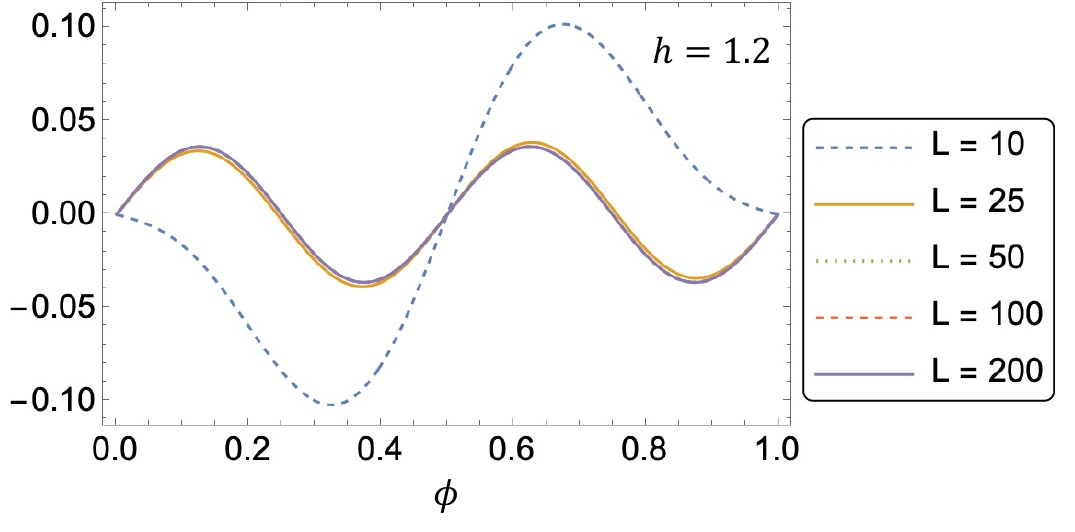}%&\includegraphics[width=60mm]{I_vs_mu_L_var_mu_var_t_1_Delta_1_phi_0.4_PBC.pdf}
    \end{tabular}
    \caption{ Current versus $\phi$ for different system size $L$. Quantum ring configuration in panels (a,~b,~c); rf-SQUID in panels (d,~e,~f). $h = 0.4$ in panels (a,~d), $h=1$ in panels (b,~e), $h=1.2$ in panels (c,~f). $\Delta = J = 1$.}
    % {[Correggere PBC con quantum ring e OBCcon rf-SQUID. Metti $\phi$ sull'asse orizzontale.]}}
    \label{fig:varie}
\end{figure*}
\end{widetext}
In this section, we focus on the clean case. Let us start considering the numerically evaluated ground-state current versus $\phi$ (see some examples of it in Fig.~\ref{fig:varie}). Let us emphasize that the current attains a nonvanishing limit for increasing $L$ only when the superconducting coupling $\Delta$ is nonvanishing. In absence of superconductivity ($\Delta=0$) the current is vanishing in the thermodynamic limit (see Appendix~\ref{normal:app} -- {the vanishing occurs in the sum over occupied states}). So the ground-state current is a supercurrent due to the presence of a Cooper-pair condensate. From now on we will fix $\Delta=J=1$.

In Fig.~\ref{fig:varie}(a,b,c) we show the ground-state current versus $\phi$ for the quantum ring and consider different values of $h$, while in Fig.~\ref{fig:varie}(d,e,f) we show the same for the rf-SQUID. We see that the curves for the quantum ring are periodic with period $\Phi_0$ (we can even show it analytically -- see Sec.~\ref{analy:sec}), while those for the rf-SQUID are periodic with period $\Phi_0/2$. {The association between configuration and periodicity is a robust feature that we observe also in the disordered case [compare with Fig.~\ref{I_phi_dirty:fig}].}

\begin{figure}
    \centering
    \begin{tabular}{c}
      (a)\\
      \includegraphics[width=80mm]{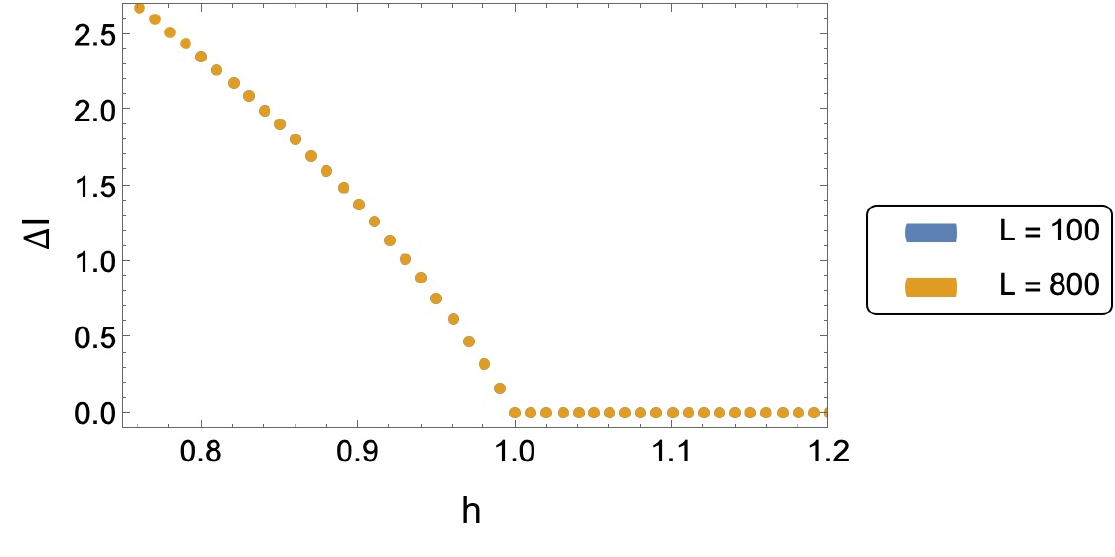}\\
      (b)\\
      \includegraphics[width=80mm]{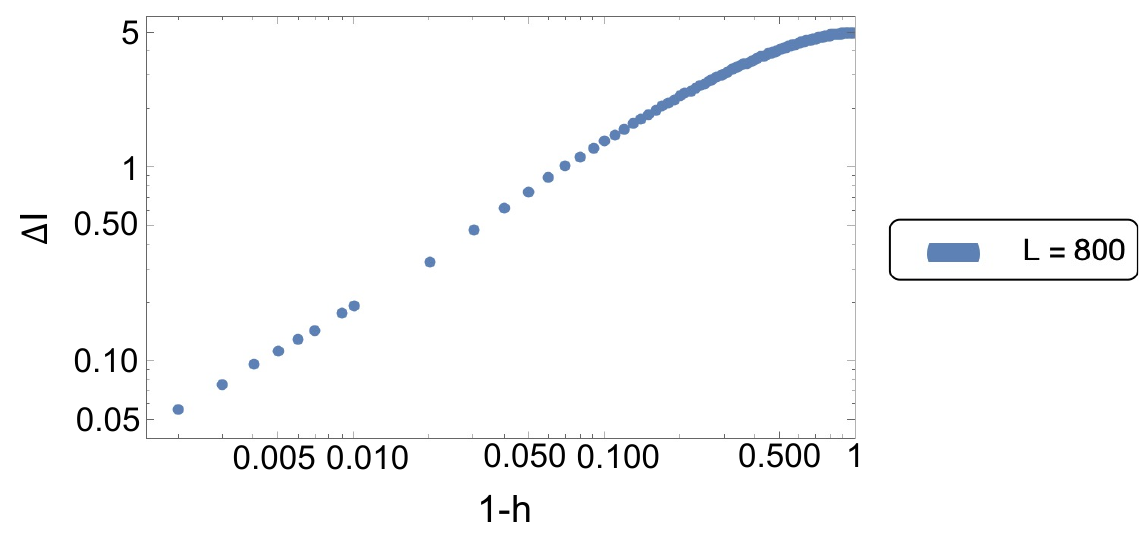}
    \end{tabular}
    \caption{(a) Jump $\Delta I$ [Eq.~\eqref{OK:eqn}] in the current versus $h$. (b) Log-Log plot of $\Delta I$ versus $1-h$. $\epsilon$ is small enough so that convergence is attained.} 
    %{[Togli OBC. Nel grafico sotto taglia a sinistra l'asse x a $h=0.0015$.]}
    \label{jump_vs_h:fig}
\end{figure}

In the topological phase there are the jumps at $\phi=1/4+n/2$ discussed in Sec.~\ref{cucud:sec}. We plot the jump $\Delta I$ versus $h$ in Fig.~\ref{jump_vs_h:fig}, and see that it shows a behavior consistent with a linear dependence on $1-h$ [Fig.~\ref{jump_vs_h:fig}(b)]. Outside the topological phase, instead, the rf-SQUID current is smooth and the $\Phi_0/2$ periodicity appears in the limit of large $L$, similarly to a symmetry-breaking effect [Fig.~\ref{fig:varie}(e)], and this is true whatever the value of $h$. 

%Therefore we see that different configurations give rise to different periodicity of the supercurrent, and both periodicities are physically meaningful, because the system generates a current in order to partially screen the external flux and  get an integer number of flux quanta $\Phi_0/2$~\cite{PhysRevLett.7.43,PhysRevLett.7.51,tinkham}. Our current has in both cases the right periodicity to get this effect, but we cannot see it in our model because we neglect the flux generated by the current.
%

%
% We report some results of the current versus $\phi$ in Fig.~\ref{I_phi_dirty:fig}, for different values of $h$ and of the boundary conditions. We see that also in the disordered case the current for the PBC case has periodicity 1 in $\phi$ (top row), and periodicity $1/2$ for OBC (bottom row), exactly as in the case of the clean system (we show here only some values of $h$, but this is true for any $h$). 
%..................................................................................................................................................................................................%

%
%
%...................................................................................................................................................................................................%
\subsection{Divergence of the current derivative in the chemical potential at the critical point}
\subsubsection{Numerical findings}
\begin{figure}
     \centering
%     \begin{subfigure}
%         \centering
     \begin{tabular}{c}
     (a)\\
        \hspace{0cm} \includegraphics[width=80mm]{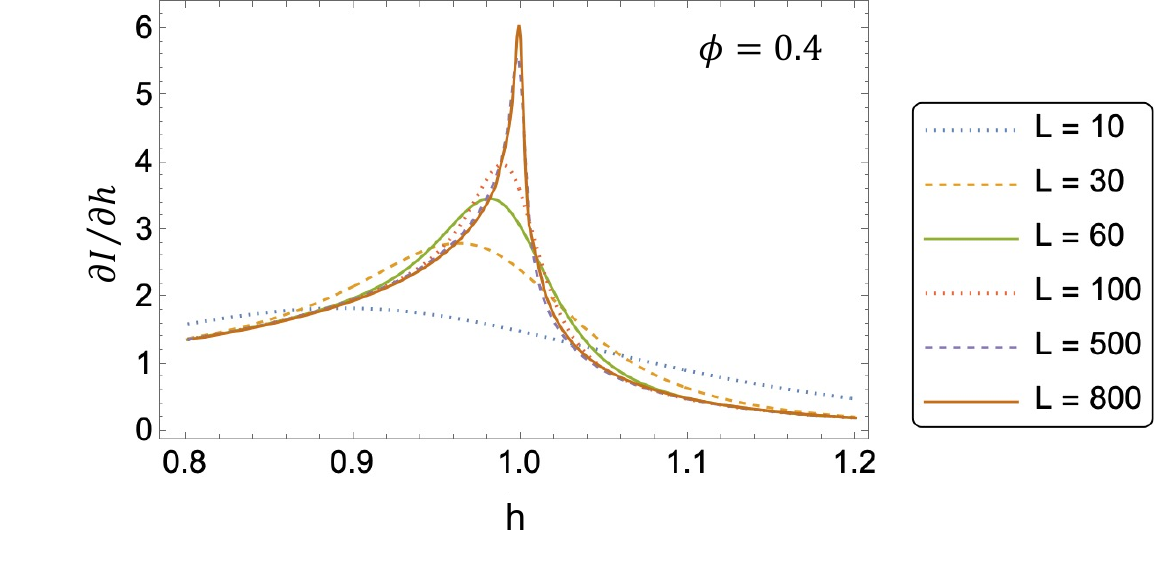}\\
%         \caption{}
%         \label{fig:deriv_I_vs_mu_L_var_mu_var_t_1_Delta_1_phi_0.8_OBC}
%     \end{subfigure}
%     \hfill
%     \begin{subfigure}
%         \centering
(b)\\
         \includegraphics[width=80mm]{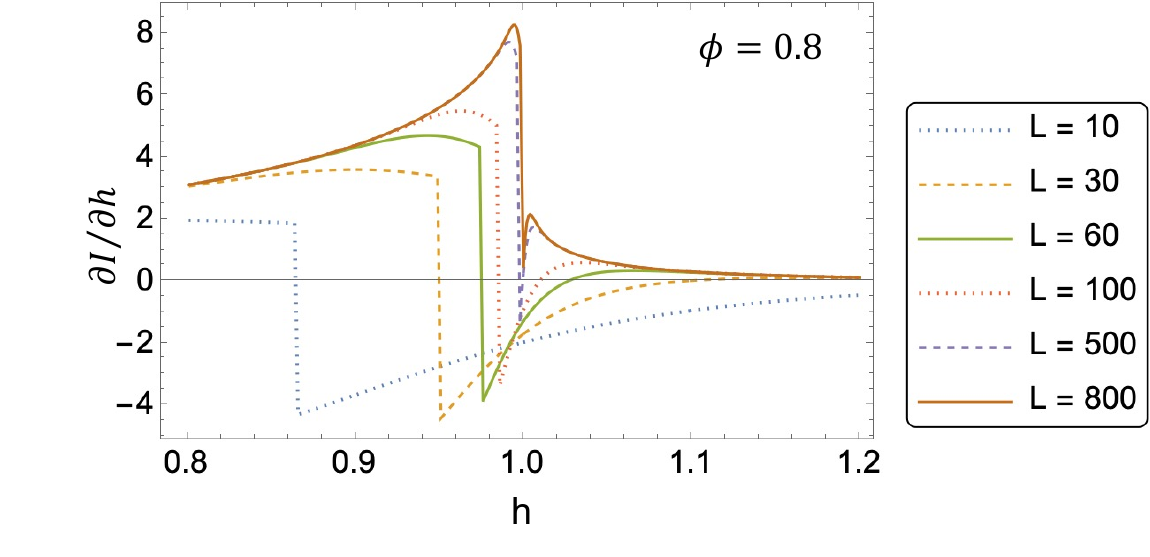}\\
         (c)\\
         \includegraphics[width=80mm]{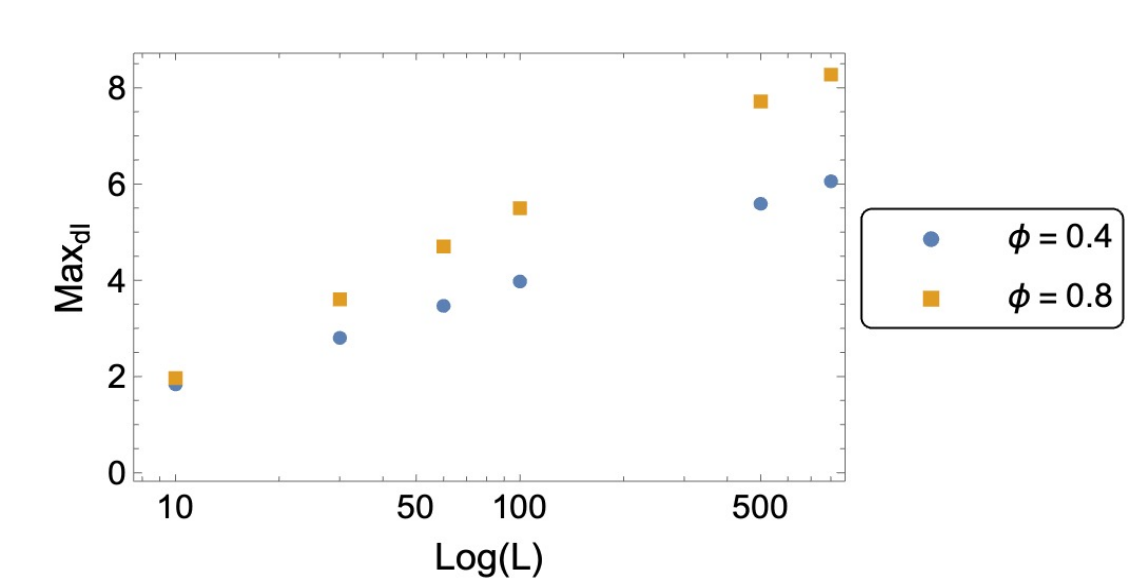}\\
     \end{tabular}
         \caption{%(a) Derivative of the current as a function of chemical potential $\mu $ ($\ud I\equiv  \frac{\partial I}{\partial h}$), at $\Phi/\Phi_0$ = 0.8$, \Delta = 1$ and OBC, for different system size $L$. (b) 
         (a,b) $\partial I/\partial h$ versus $h=\mu/2$, in the rf-SQUID at two values of $\phi$, for different system sizes $L$. Maximum over $h$ of $\ud I \equiv \partial I/\partial h$ versus $\log_{10}(L)$, for two values of $\phi$ in the rf-SQUID. Notice the linear dependence, that corresponds to the logarithmic divergence $\max \ud I\propto \log L$. (c) }
         %{[Togli OBC e sostituisci $dI$ con $\partial I/\partial h$. Sostituisci $\Phi/\Phi_0$ con $\phi$.]}}
         \label{fig:derivate}
%     \end{subfigure}
\end{figure}

Let us numerically evaluate the derivative in $\mu$ of the current, $\partial I/\partial\mu = \frac{1}{2}\partial I /\partial h$. If there is a finite-size discontinuity point, we evaluate the right and the left derivatives. In Fig.~\ref{fig:derivate}(a,b)~\cite{panota} we show two examples of $\partial I/\partial h$ versus $h$. We see a peak near the transition point; For increasing system size the position in $h$ of the peak tends towards the critical point, while its height diverges logarithmically with $L$. We can see this divergence in Fig.~\ref{fig:derivate}(c) where we plot versus $L$ the maximum of the current, taking a logarithmic scale on the horizontal axis, and get straight lines. 

%So in the thermodynamic limit the current derivative develops a divergence at the critical point. ()

The logarithmic divergence of Fig.~\ref{fig:derivate}(c) is quite robust, and we have observed it whatever value of $\phi$ {(provided the current is nonvanishing), and in both} configurations. It is a consequence of the fact that the transition is second order, so second-order derivatives of the ground-state energy -- like $\partial I/\partial\mu$ -- show singularities. It is already known that $\chi=\partial^2\mathcal{E}/\partial h^2$ logarithmically diverges at the critical point, for $\phi=0,\,1/2$~\cite{Sachdev}. $\chi$ can be interpreted as the magnetic susceptibility of model in {the Ising-model representation}, but no similar findings where known until now for quantities simple to measure in the Kitaev model.

% Here we fill this gap, and find the physical result that one can see the singular behavior at the critical point by measuring a supercurrent and performing its derivative in the chemical potential, a finding of interest for experiments, and a way to locate the topological transition and indirectly probe the presence of the topological phase and the related Majorana fermions. 
In the next section we are going to see that {in the quantum-ring configuration} and small $|\sin(2\pi\phi)|$ {one can even analytically predict that $\partial I/\partial\mu$ diverges as $\log L$ for finite size at the transition point, and as $\log|J-h|$ near the transition point in the thermodynamic limit.}
%To further study the case of OBC, we compute the derivative of the current as $\mu$ varies, for different system sizes. In Fig.\ref{fig:deriv_I_vs_mu_L_var_mu_var_t_1_Delta_1_phi_0.4_OBC} we illustrate the derivative of the current calculated at $\Phi/\Phi_0 = 0.8$ where $I$ shows singularities in its behavior. Here, a maximum of $dI\equiv \frac{\partial I}{\partial h}$ is visible before the discontinuity point. In Fig.\ref{fig:deriv_I_vs_mu_L_var_mu_var_t_1_Delta_1_phi_0.4_OBC} we show $\ud I$ versus $h$ at $\Phi/\Phi_0 = 0.4$, corresponding to a case where $I$ is continuous by varying the chemical potential $\mu$.
%
\\
%
%
%.....................................................................................................................................................%
\subsubsection{Analytical interpretation}\label{analy:sec}
{Let us consider Eq.~\eqref{hbt:eqn} in the quantum-ring configuration.} We consider the boundary flux term as a perturbation on the unperturbed uniform system. This approach is sensible if $2|\sin(\pi\phi)|\ll 1$ and succeeds in predicting the periodicity in $\phi$ with period 1 of the current, and the divergence of $\partial I/\partial h$ at the critical point $h=J$. {Let us write the Hamiltonian as} %(see Sec.~\ref{new:sec}). Quite nicely, modes at the spectral edges appear, that provide a contribution to the current with the correct periodicity in $\phi$ (see Sec.~\ref{modes:sec}).

%\subsection{New representation}\label{new:sec}
%
%We can analytically interpret the logarithmic divergence of the peak of the derivative of the current near the critical point, that we see in Fig.~\ref{fig:max_dI_vs_L}. 
%We focus on the case with PBC in the superconducting terms, and assume that we are in the representation where the flux acts only as a boundary term. In this representation we can write the Hamiltonian as
%
\begin{align}
   \hat{H} &= \hat{H}_0 + \hat{H}_1\quad\text{with}\\
   \hat{H}_0&\equiv 2h\sum_j\opcdag{j}\opc{j} -J\sum_j(\opcdag{j}\opc{j+1}+\text{H.~c.})\nonumber\\
   &+ \Delta \sum_j(\opcdag{j}\opcdag{j+1}+\text{H.~c.})\quad\text{and}\\
   \hat{H}_1&\equiv -J(\nep^{-i2\pi\phi}-1)\opcdag{L}\opc{1} + \text{H.~c.}\,,
\end{align}
where we have defined $h=\mu/2$. If $|\nep^{-i2\pi\phi}-1|=2|\sin(\pi\phi)||\ll 1$, we can solve $\hat{H}_0$ and treat $\hat{H}_1$ perturbatively. To solve $\hat{H}_0$ we apply the Fourier transform $\opc{j} = \frac{1}{\sqrt{L}}\sum_k\opc{k}\nep^{ikj}$ and straightforwardly write
\begin{align}\label{hoho:eqn}
    \hat{H}_0 &= 2\sum_k(h-J\cos k)\opcdag{k}\opc{k} +\Delta\sum_k \opcdag{k}\opcdag{-k}\nep^{-ik}\nonumber\\
     &+ \left(\Delta\sum_k \opc{-k}\opc{k}\nep^{ik}+{\rm H.~c.}\right)\,.
\end{align}
In order to be consistent with the periodic boundary conditions imposed on $\hat{H}_0$, the allowed values of $k$ are 
\begin{equation}
    k_n = \frac{2\pi n}{L}\, %+2\pi\frac{\phi}{L}\,,
\end{equation} 
with $n\in\{-L/2+1,\,\ldots,\,L/2\}$ integer. %Defining $k_n\equiv \frac{2\pi n}{L}$ %and $a\equiv 2\pi /L$
We can rewrite Eq.~\eqref{hoho:eqn} as
\begin{align}\label{hoho1:eqn}
%    \hat{H}_0 = 2\sum_{n}[h-J\cos (k_n+a\phi)]\opcdag{q_n}\opc{q_n} +\Delta\sum_{n} \opcdag{q_n}\opcdag{q_{-n}}\nep^{-ik_n-ia\phi} + \Delta\sum_k \opc{q_{-n}}\opc{q_n}\nep^{ik_n}\,. %+ia\phi}\,.
    \hat{H}_0 &= 2\sum_{n}[h-J\cos (k_n)]\opcdag{k_n}\opc{k_n} +\Delta\sum_{n} \opcdag{k_n}\opcdag{-k_{n}}\nep^{-ik_n}\nonumber\\
              & + \left(\Delta\sum_n \opc{-k_{n}}\opc{k_n}\nep^{ik_n}+{\rm H.~c.}\right)\,. %+ia\phi}\,.
\end{align}
Coupling each $k_n$ with the corresponding $-k_{n}$ we can rewrite this formula as
\begin{align}
%    \hat{H}_0 &= 2[h-J\cos(a\phi)]\opcdag{0}\opc{0} + 2\sum_{n>0}[h-J\cos(a\phi-k_n)] \nonumber\\
%    &+2\sum_{n>0}\left(\begin{array}{cc}\opcdag{q_n}&\opc{q_{-n}}\end{array}\right)
%      \left(\begin{array}{cc}[h-J\cos(k_n+a\phi)]&-i\Delta\sin k_n\nep^{-ia\phi}\\
%                                    i\Delta\sin k_n\nep^{ia\phi}&-[h-J\cos(k_n-a\phi)]\end{array}\right)
%                            \left(\begin{array}{c}\opc{q_n}\\\opcdag{q_{-n}}\end{array}\right)\,,
    \hspace{-0.2cm}{\footnotesize \hat{H}_0 \hspace{-0.1cm}=2\!\sum_{n>0}\hspace{-0.1cm}\left(\begin{array}{cc}\opcdag{k_n}&\!\opc{-k_{n}}\end{array}\right)\hspace{-0.1cm}
      \left(\begin{array}{cc}[h-J\cos(k_n)]&\hspace{-0.2cm}-i\Delta\sin k_n\\
                                    i\Delta\sin k_n&\hspace{-0.2cm}-[h-J\cos(k_n)]\end{array}\right)\hspace{-0.1cm}
                            \left(\begin{array}{c}\opc{k_n}\\\opcdag{-k_{n}}\end{array}\right).}
\end{align}
%
%where $a=2\pi/L$ and $\phi=\Phi/\Phi_0$. The precise value of the $k$s depends on the choice of the boundary conditions, but are generically of the form 
%\begin{equation}\label{kakka:eqn}
%    k=k_0+\frac{2\pi n}{L}\in [0,\pi]\,.
%\end{equation}
%
%Neglecting the case of a $k=0$ that appears for $k_0=0$ and undergoes a special treatment~\cite{glen}, we see that there is a $k*$ that is the nearest possible to 0 and has asymptotically the form $k^*\sim b/L$ (if it has not this form, one could not fit $L/2$ equally spaced values of $k$ in the interval $[0,\pi]$)
%Notice that the flux acts only on the normal terms and its effect is to shift the value of $k$ in the cosine. 
With an analysis strictly similar to the one in~\cite{glen}, the ground state %energy (the one of the state without quasiparticle excitations) can be shown to be
of this Hamiltonian can be shown to be
\begin{equation}\label{stato:eqn}
    \ket{\psi} = \prod_{n>0}(v_{n} + u_{n}\opcdag{k_n}\opcdag{-k_{n}})\ket{0}_c\,,
\end{equation}
where $\ket{0}_c$ is the vacuum of the $\opc{k_n}$ operators, and
\begin{align}\label{uva:eqn}
%  u_n&=i\sin\left(\frac{\theta_n}{2}\right)\nep^{ia\phi/2}\nonumber\\
%  v_n&=\cos\left(\frac{\theta_n}{2}\right)\nep^{-ia\phi/2}\quad\text{with}\quad \tan\theta_n = \frac{\Delta\sin k_n}{h-J\cos(a\phi)\cos k_n}\,.
  u_n&=i\sin\left(\frac{\theta_n}{2}\right)\nonumber\\
  v_n&=\cos\left(\frac{\theta_n}{2}\right)\quad\text{with}\quad \tan\theta_n = \frac{\Delta\sin k_n}{h-J\cos(a\phi)\cos k_n}\,.
\end{align}
%
%There is an important point to notice, the fact that $q_n$ is coupled with $q_{-n}$ but $q_n\neq -q_{-n}$. This is at variance with what happens in absence of magnetic field (or in the bad representation discussed in Sec.~\ref{bad:sec}). The point is that the magnetic field breaks the time-inversion symmetry, as Procolo noticed some days ago.

With the ground state evaluated in this approximation  (Eqs.~\eqref{stato:eqn},~\eqref{uva:eqn}) we can compute the expectation of the current operator, that in this representation is~\cite{supp,thouless}
\begin{equation}
    \hat{I} = 2\pi\left.\frac{c}{\Phi_0}\frac{\partial\hat{H}}{\partial \phi}\right|_{\phi=0} = iJ\left(\nep^{2\pi i \phi}\opcdag{L}\opc{1}-\nep^{-2\pi i \phi}\opcdag{1}\opc{L}\right)\,,
\end{equation}
%\nep^{i2\pi\phi}\nep^{-i2\pi\phi}
where we have used that we are assuming $2\pi c/\Phi_0 = 1$. Applying the Fourier transform we get
\begin{equation}
    \hat{I}=i\frac{J}{L}\nep^{2\pi i \phi}\sum_{n,\,n'}\nep^{-ik_n L+ik_n'}\,\opcdag{k_n}\opc{k_{n'}}+\text{H.~c.}\,.
\end{equation}
Evaluating the expectation of this operator on the approximate ground state provided by Eqs.~\eqref{stato:eqn},~\eqref{uva:eqn} we get (see Appendix~\ref{alg:app}) for details
\begin{align}\label{i_ota:eqn}
   &\langle I\rangle \simeq -\Im{\frac{2J}{L}\nep^{2\pi i\phi}\sum_{n\geq 0}\cos(k_n)}\nonumber\\
   &+\Im{\frac{J}{L}\nep^{2\pi i\phi}\sum_{n\geq 0}\frac{(h-J\cos k_n)\cos k_n}{\sqrt{(h-J\cos k_n)^2+J^2\sin^2 k_n}}}\,.
\end{align}
Moving to the thermodynamic limit, the sums become integrals ($\frac{1}{L}\sum_n(\ldots)\to\frac{1}{2\pi}\int_0^\pi(\ldots)\ud k$). In this way one easily sees that in this limit the first contribution vanishes, and the current becomes %, and neglected a term that does not depend on $h$. In this limit we get %simply $\langle I\rangle = 0$. One should go to further orders in the expansion in $\Delta(1-\cos(2\pi\phi))$ but I do not know how to do it.
\begin{equation}
    \langle I\rangle \simeq
    -\sin(2\pi\phi)\frac{J}{2\pi}\int_{0}^\pi\frac{(h-J\cos k)\cos k}{\sqrt{h^2+J^2-2hJ\cos k}}\ud k\,. %and $aL =2\pi$
\end{equation}
So the current that we have evaluated is periodic  in $\phi$ with period 1. We emphasize that this formula is valid for $|\nep^{2\pi i \phi}-1|\ll 1$, that's to say $2\pi|\phi-n|\ll 1$ for some $n$ integer. So the formula is valid near $\phi=n$ and correctly predicts that the current is odd in $(\phi-n)$ and vanishes at $\phi=n$, in agreement with the numerics (see bottom row of Fig.~\ref{fig:varie}). 

Let us now evaluate the derivative in $\mu$ of the current. From the definition it is given by $ \frac{\partial \langle I\rangle}{\partial \mu}=\frac{1}{2} \frac{\partial \langle I\rangle}{\partial h}$. For finite size we get
\begin{equation}\label{ifino:eqn}
    \frac{\partial \langle I\rangle}{\partial \mu}=-\frac 12\sin(2\pi\phi)\frac{1}{L}\sum_{n > 0}\frac{J(J-h\cos k_n)\cos k_n}{\left[h^2+J^2-2hJ\cos k_n\right]^{3/2}}\,,
\end{equation}
where the terms for $k_n=0$ and $k_n=\pi$ vanish in the derivative because they do not depend on $h$. In the thermodynamic limit we get
\begin{equation}\label{ico:eqn}
    \frac{\partial \langle I\rangle}{\partial \mu}=-\sin(2\pi\phi)\frac{1}{4\pi}\int_{0}^\pi\frac{J(J-h\cos k)\cos k}{\left[h^2+J^2-2hJ\cos k\right]^{3/2}}\ud k\,.
\end{equation}
When $h=J$ this integral reduces to
\begin{equation}\label{ico1:eqn}
    \left.\frac{\partial \langle I\rangle}{\partial \mu}\right|_{h=1}=-\sin(2\pi\phi)\frac{1}{32\pi}\int_{0}^\pi\frac{\cos k}{\sin(k/2)}\ud k\,.
\end{equation}
One can clearly see that it shows logarithmic divergences for $k=0$ and $k=\pi$. At finite size $L$, the integral is approximated by the finite sum Eq.~\eqref{ifino:eqn}, whose extrema are $k_{-}=2\pi/L$ and $k_{+}=\pi-2\pi/L$. If $h=J$ and $L$ is large, we approximate the sum with the integral in Eq.~\eqref{ico:eqn} with extrema $k_-$ and $k_+$ (instead as $0$ and $\pi$). That provides a contribution that diverges logarithmically in $L$. We have already numerically observed this divergence in Fig.~\ref{fig:derivate}(a). Moreover, the integral in Eq.~\eqref{ico:eqn} can be easily shown to logarithmically diverge as $\sim\log(|J-h|)$ for $h\to J$.
%--------------------------------------------------------------------------------------------------------------------------------------------------------------------------------------------------%
\section{Conclusion}\label{conc:sec}
In conclusion we have studied a Kitaev chain with a magnetic flux, considering different configurations: The case of a rf-SQUID, where a weak link is present, and the case of a fully superconducting ring. We have first focused on the disordered case. 

The system in the ground state  displays a supercurrent induced by the flux, and considering its dependence on the flux, we have seen that it is periodic, with a periodicity depending on the chosen configuration. The period is $\Phi_0/2$ in the rf-SQUID for large system size, and $\Phi_0$ in the quantum ring, where $\Phi_0 = 2\pi\hbar c/e$ is the flux quantum. Both periodicities are consistent with the fact that the phase of the condensate wavefunction must be single valued, and the external flux plus the one induced by the current must be an integer multiple of $\Phi_0/2$~\cite{PhysRevLett.7.43,PhysRevLett.7.51,tinkham}. The association between configuration and periodicity is very robust and we observe it numerically for all the considered parameters, both for the disordered and the clean case.

In the rf-SQUID case, when the system is in the topological phase, the current shows some jumps of finite amplitude. These jumps in the thermodynamic limit appear at fixed values of the flux $\Phi^*=\Phi_0/4+n\Phi_0/2$, as it has been shown in~\cite{Nava_2017} for the clean case. Here, with a simple perturbative argument, we show that they are a consequence of the existence of the Majorana fermions and the time-reversal symmetry in the open chain: Adding the weak link, the double-degeneracy of the ground state due to the Majoranas is broken, and a topological gap opens everywhere but at $\Phi=\Phi^*$. Here the ground-state energy shows some cusps, leading to the discontinuities in the current.

We thereby generalize the argument of~\cite{Nava_2017}, and show that the current jumps and their position in $\Phi$  in the thermodynamic limit only depend on the presence of the Majorana modes, without direct reference to the specific form of the disorder (or the absence thereof). Both the current jump for $\Phi=\Phi^*$ and the topological spectral gap for $\Phi\neq \Phi^*$ are useful to probe the topological phase in the disordered case, where quantities appropriate for the clean case do not work so well.

In the disordered case we  use the scaling of the inverse participation ratio to show that the quasiparticle excitations are Anderson localized. {For $\Delta = J$, where the quasiparticles without disorder display a flat band, even the tiniest disorder is enough to induce localization with a finite localization length.} %, also for very small disorder amplitudes. 
So, while the condensate is delocalized and carries supercurrent also in the disordered case, the excitations are localized and cannot carry any current. This implies an interesting consequence: At finite temperature -- as long as the quasiparticle description is meaningful -- the system carries only supercurrent without resistive contributions. Moreover, if the superconducting coupling {lies at $\Delta = J$}, the localization length is finite even for very small disorder strength, at variance with what happens in the normal case.

For the clean case we have seen, both numerically and analytically, that the derivative of the supercurrent with respect to the chemical potential diverges at the critical point of the topological-to-trivial transition. This is consistent with the fact that the transition is known to be second order, and provides a way -- independent of the boundary conditions and of the existence of Majorana boundary fermions in the topological phase -- for probing the topological-to-trivial transition.

%This argument relies only on the existence of the boundary Majorana fermions and on the time-reversal invariance of the open chain, and applies also to the case of a chain with disorder (dirty case). Indeed, we numerically see that the jump in the current vanishes exactly where analytical arguments based on the renormalization group predict the topological-to-trivial phase transition~\cite{fisher_prb95}. So, for OBC, the presence of the jump is a way to probe the existence of Majorana fermions also in the dirty case.

Future perspectives include the application of these methods (especially the research of divergences of the current derivative at the topological-to-trivial transition) to the case of more realistic models of topological superconductors~\cite{Marra_2016}, and the study of the relation between work statistics~\cite{PhysRevB.91.205136,Russomanno_2015} and topological properties in the case of a time-varying flux. {Moreover, if our predictions are confirmed to apply also to more realistic models of topological superconductors~\cite{Marra_2016}, they could be experimentally tested using the scanning SQUID microscopy~\cite{scasq}. (With this method one measures the circulating current in the quantum-ring or rf-SQUID configurations threaded by a flux, by probing the contribution to the magnetic field generated by the current.)} %{\bf [Qua il Referee ci chiede di ampliare. Bisogna chiedere a Procolo e Gabriele che sono degli esperti.]}
\acknowledgements{A.~R. acknowledges interesting discussions with V.~Russomanno and A.~Silva. M.~M. acknowledges R.~Capecelatro for fruitful discussions. {We acknowledge financial support and computational resources from MUR, PON ``Ricerca e Innovazione 2014-2020'', under Grant No. ``PIR01\_~00011 -- (I.Bi.S.Co.)''. A.~R. acknowledges financial support from PNRR MUR project PE0000023-NQSTI. P.~L. acknowledges financial support from  PNRR MUR project CN\_~00000013-ICSC,  as well as from the project QuantERA II Programme STAQS project that has received funding from the European Union's Horizon 2020 research and innovation programme under Grant Agreement No 101017733. }}
\appendix
%-------------------------------------------------------------------------------------------------------------------------------------------------------------------------------------------------%
\section{Some algebra}\label{alg:app}
Eq.~\eqref{i_ota:eqn} is obtained in the following way
\begin{widetext}
\begin{align}
    \langle I\rangle &\simeq \Re{i\frac{J}{L}\nep^{2\pi i\phi}\sum_{n\geq 0}\left(\nep^{ik_n(1-L)}\langle\opcdag{k_n}\opc{k_n}\rangle+\nep^{-ik_n(1-L)}-\nep^{-ik_n(1-L)}\langle\opc{-k_{n}}\opcdag{-k_{n}}\rangle\right)}\nonumber\\
    &=\Re{i\frac{J}{L}\nep^{2\pi i\phi}\sum_{n\geq0}\left(\nep^{ik_n}\sin^2\left(\frac{\theta_n}{2}\right)+\nep^{-ik_n}\sin^2\left(\frac{\theta_n}{2}\right)\right)}\nonumber\\
    &=\Re{i\frac{2J}{L}\nep^{2\pi i\phi}\sum_{n\geq 0}\cos(k_n)\sin^2\left(\frac{\theta_n}{2}\right)}\nonumber\\
    &= -\Im{\frac{2J}{L}\nep^{2\pi i\phi}\sum_{n\geq 0}\cos(k_n)}+\Im{\frac{J}{L}\nep^{2\pi i\phi}\sum_{n\geq 0}\frac{(h-J\cos k_n)\cos k_n}{\sqrt{(h-J\cos k_n)^2+J^2\sin^2 k_n}}}\,,
\end{align}
\end{widetext}
where we have exploited that $k_n L = 2n\pi$. 
%------------------------------------------------------------------------------------------------------------------------------------------------------------------------------------------------%
\section{Current discontinuities at the topological transition in the clean case}\label{disco:sec}
\begin{figure}
     \centering
     \begin{tabular}{cc}
         %\centering
%       \hspace{0cm}  \\
%         \caption{}
%         \label{fig:I_vs_mu_L_var_mu_var_t_1_Delta_1_phi_0.4_PBC}
%     \end{subfigure}
%     \hfill
%     \begin{subfigure}
%         \centering
(a)\\
        \includegraphics[width=80mm]{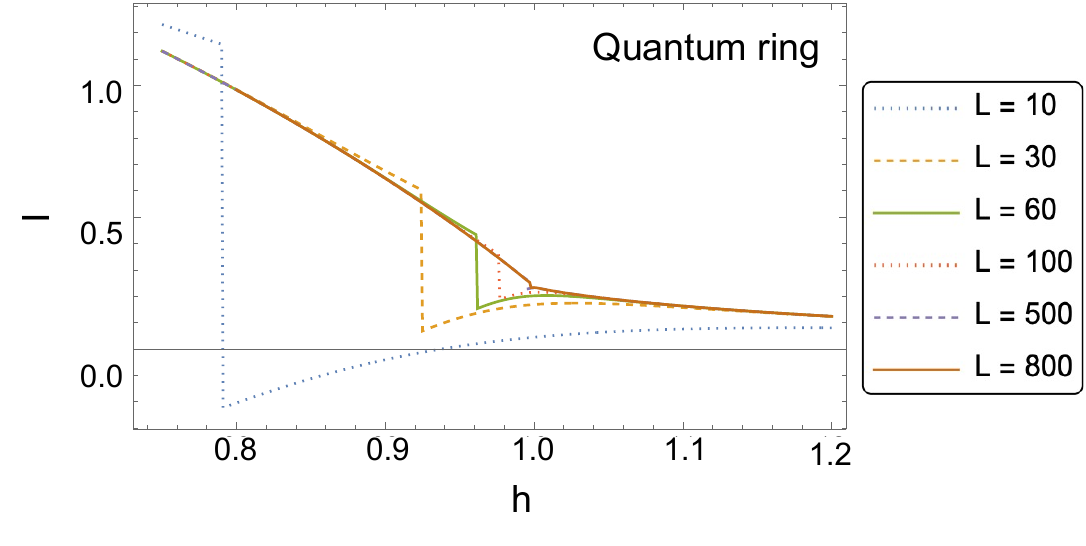}\\
        (b)\\
         \includegraphics[width=80mm]{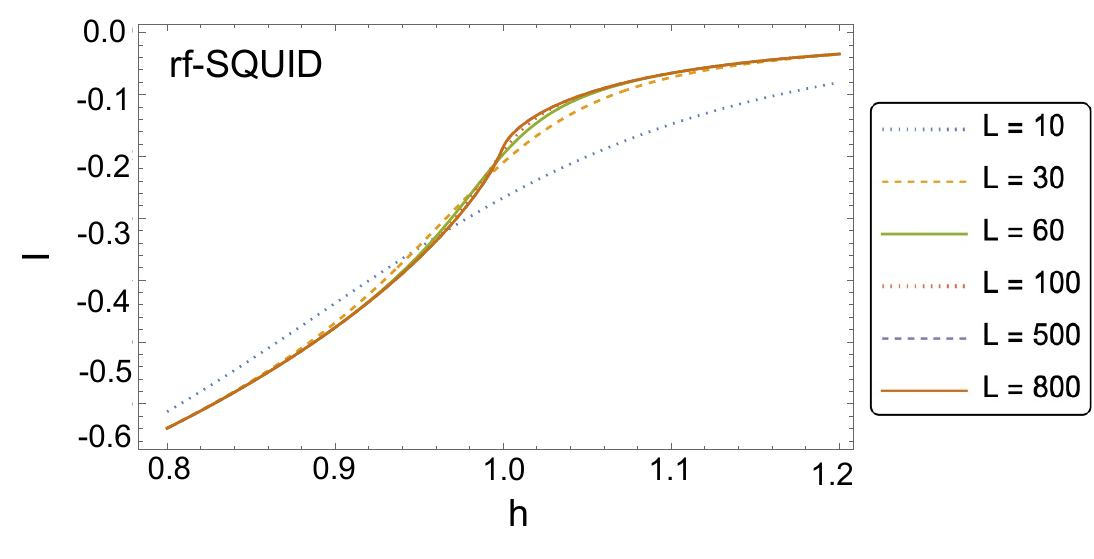}
     \end{tabular}
         \caption{Current versus $h$ for $\phi=0.4$ in the quantum ring (panel a), and rf-SQUID (panel b). Notice the discontinuity near the critical point that shrinks with increasing system size.}
         % {[Sostituisci OBC con rf-SQUID e PBC con quantum ring. Togli $\Phi/\Phi_0=0.4$ dalla caption.]}}
         \label{fig:I_vs_mu_L_var_mu_var_t_1_Delta_1_phi_0.8_PBC}
\end{figure}

\begin{figure}
    \centering
    \begin{tabular}{c}
     \hspace{0cm}\includegraphics[width=60mm]{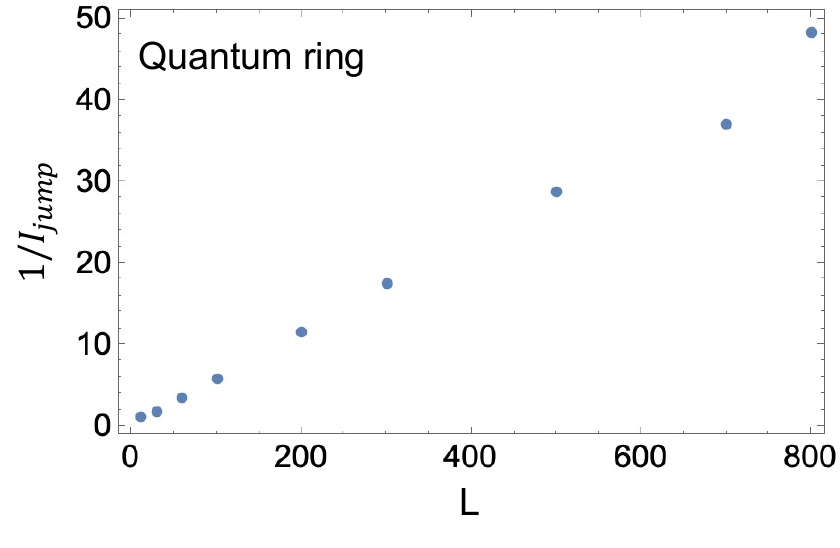}
    \end{tabular}
    \caption{Inverse of the discontinuity in the current versus $L$, in the quantum ring for $\phi = 0.4$.}
    % {[Sostituisci PBC con quantum ring.  Togli $\Phi/\Phi_0=0.4$ dalla caption.]}
    \label{fig:IJump_vs_L_phi_0.4_PBC}
\end{figure}

Let us fix $\Delta = J = 1$. Let us analyze the current versus $h$ {(see Fig.~\ref{fig:I_vs_mu_L_var_mu_var_t_1_Delta_1_phi_0.8_PBC})}. The first thing we notice is that in many cases there is a discontinuity of the current near the transition point $h=J$. This discontinuity is different than the jumps discussed in Sec.~\ref{cucud:sec}, because its height decreases with increasing system size as $\sim 1/L$ and so tends vanish in the thermodynamic limit (see Fig.~\ref{fig:IJump_vs_L_phi_0.4_PBC}). In this limit, therefore, the current becomes continuous in $h$ at the transition. This is consistent with the transition being second order, so in the thermodynamic limit singularities can appear at the transition point only in the second derivatives of the ground-state energy.

%We also analyzed the current behavior as a function of the chemical potential $\mu$, at a fixed value of $\Phi/\Phi_{0}$. With PBC, $I$ shows some singularities for both the chosen values of $\Phi/\Phi_{0} = 0,4$ and $\Phi/\Phi_{0} = 0.8$ , as illustrated in Fig.~\ref{fig:I_vs_mu_L_var_mu_var_t_1_Delta_1_phi_0.8_PBC}.  \\
%
%\\
%These current singularities also occur at some values of $\Phi/\Phi_{0}$ with OBC [see Fig.\ref{fig:I_vs_mu_L_var_mu_var_t_1_Delta_1_phi_0.8_OBC}(left)]. However, in this case, the current discontinuity seems to depend on the particular value of $\Phi/\Phi_{0}$, and the current can also appear continuous as the chemical potential varies, as shown in Fig.\ref{fig:I_vs_mu_L_var_mu_var_t_1_Delta_1_phi_0.8_OBC}(right) at $\Phi/\Phi_{0} = 0,4$. \\
%\begin{figure}
     %\centering
%     \begin{subfigure}
%         \centering
%     \begin{tabular}{c}
%       \hspace{0cm} \\
%         \caption{}
%         \label{fig:I_vs_mu_L_var_mu_var_t_1_Delta_1_phi_0.4_OBC}
%     \end{subfigure}
%     \hfill
%     \begin{subfigure}
%         \centering
%         (a)\\
%         \includegraphics[width=80mm]{{imgs_6_06/I_vs_h_0.8_PBC}.pdf}\\
%         (b)\\
%         \includegraphics[width=80mm]{{imgs_6_06/I_vs_h_0.8_OBC}.pdf}
%
%     \end{tabular}
%         \caption{Current versus $h$ for $\phi=0.8$ in the quantum ring (panel a), and in the rf-SQUID (panel b). Notice the discontinuity near the critical point that shrinks with increasing system size.} 
         %{[Sostituisci OBC con rf-SQUID e PBC con quantum ring.  Togli $\Phi/\Phi_0=0.8$ dalla caption.]}}
%         \label{fig:I_vs_mu_L_var_mu_var_t_1_Delta_1_phi_0.8_OBC}
%
%\end{figure}
%
\begin{figure}
     \centering
     \begin{tabular}{cc}
%         \centering
(a)\\
        \hspace{0cm} \includegraphics[width=82mm]{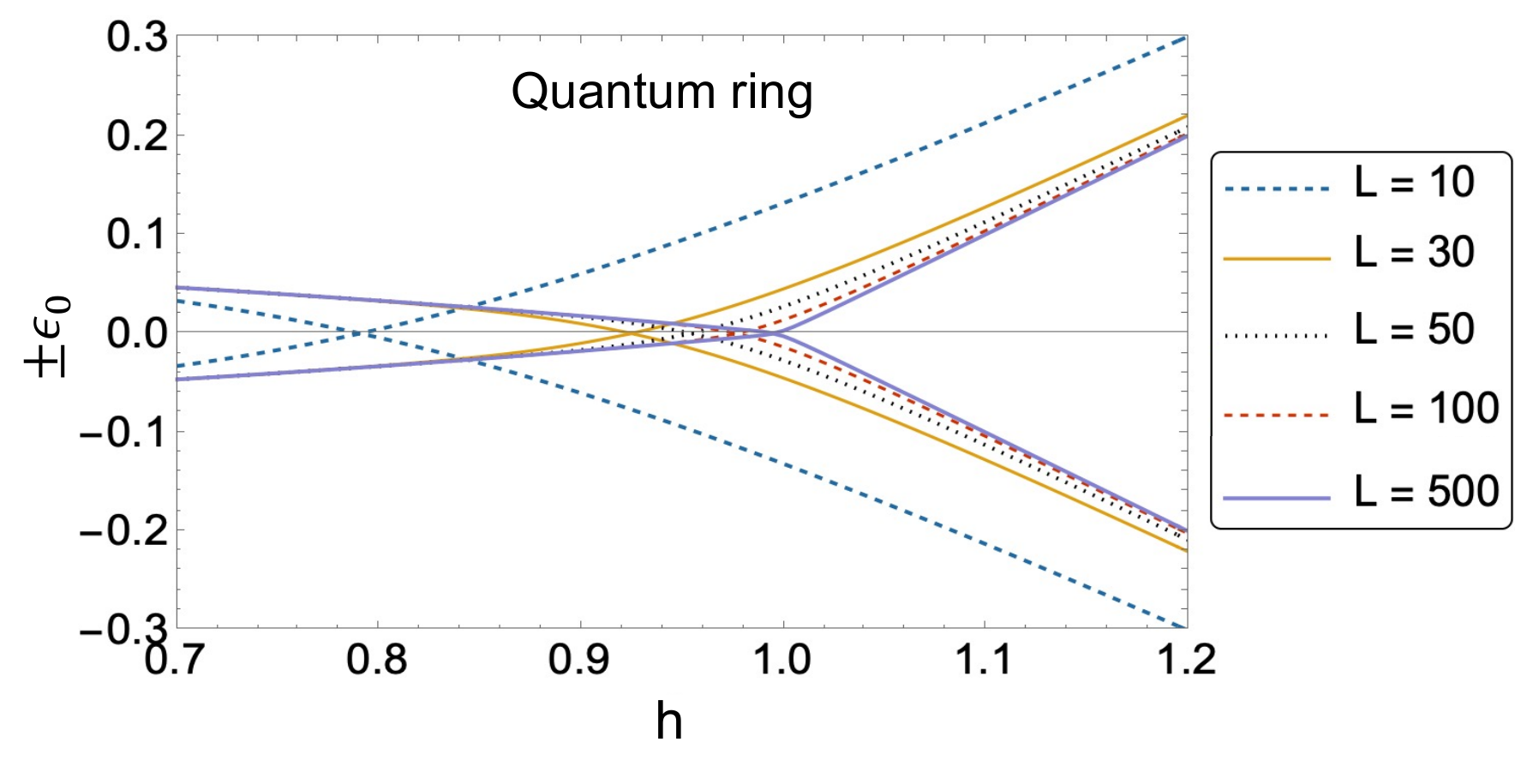}\\
%         \caption{}
%         \label{fig:evalues_vs_mu_phi_0.4_PBC}
%%     \end{subfigure}
%     \hfill
%     \begin{subfigure}
%         \centering
(b)\\
         \includegraphics[width=82mm]{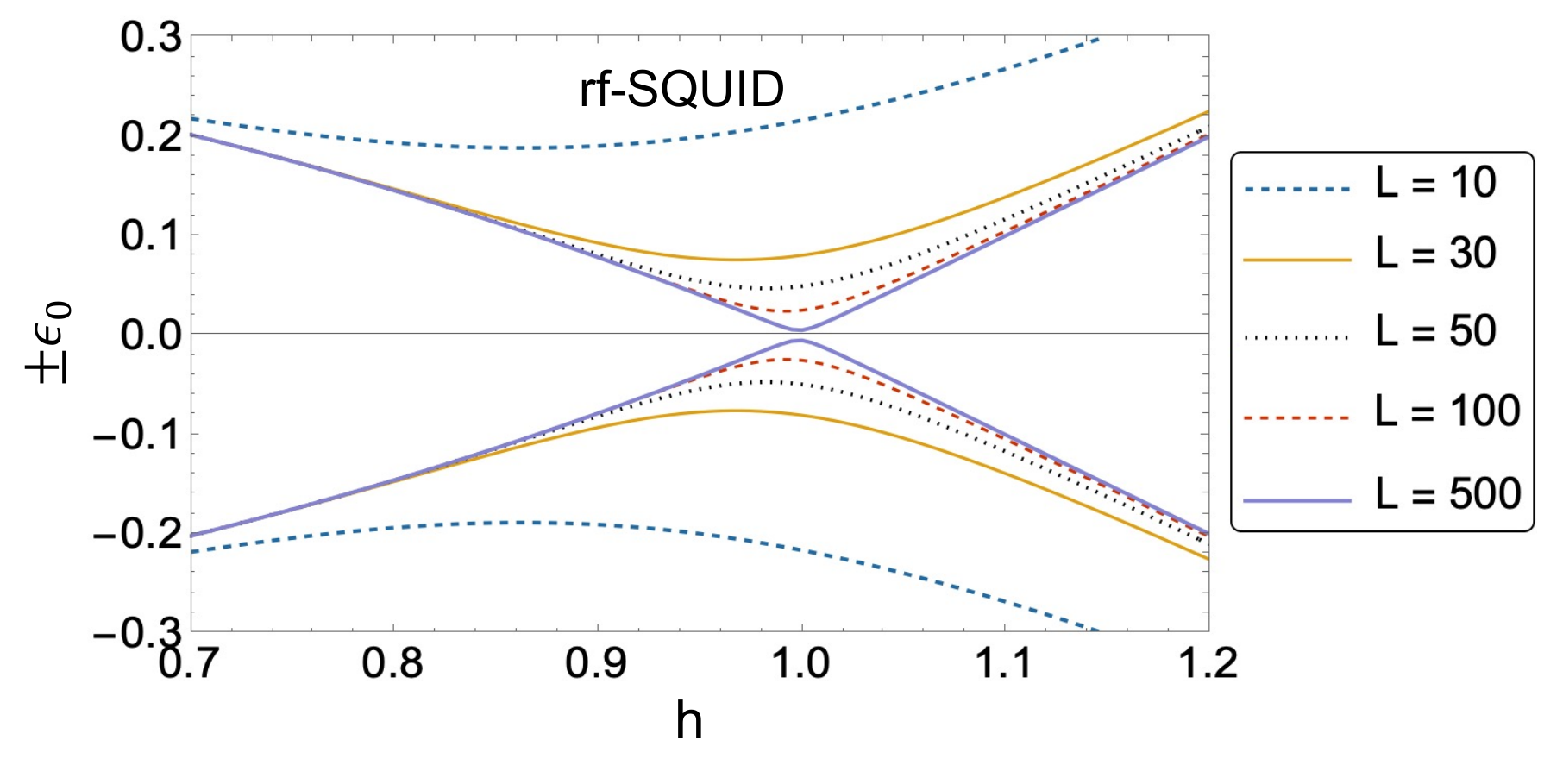}
     \end{tabular}
         \caption{(a) Quasiparticle energies $\pm\epsilon_0$ versus $h$ in the quantum ring for $\phi = 0.4$ , and different system sizes $L$. Notice the crossing point corresponding to the discontinuity in the current [Fig.~\ref{fig:I_vs_mu_L_var_mu_var_t_1_Delta_1_phi_0.8_PBC}(a)]. (b) The same in the rf-SQUID. In this case, the crossing of levels no longer occurs, and correspondingly there is no current discontinuity  [Fig.~\ref{fig:I_vs_mu_L_var_mu_var_t_1_Delta_1_phi_0.8_PBC}(b)]. }
         %{[Sostituisci OBC con rf-SQUID e PBC con quantum ring]}}
         \label{fig:evalues_vs_mu_phi_0.4_OBC}
%     \end{subfigure}
\end{figure}

This finite-size discontinuity in $h$ of the current is related to the crossing of the $\pm\epsilon_0$, as we see in Fig.~\ref{fig:evalues_vs_mu_phi_0.4_OBC}(a), where the crossing point tends towards the transition point $h=J$ as the size $L$ increases. We plot also a case without discontinuity [Fig.~\ref{fig:evalues_vs_mu_phi_0.4_OBC}(b)] and see that the levels show an anticrossing with a gap that tends to vanish in the large-size limit, as common for a second order quantum phase transition~\cite{Sachdev}. Let us again emphasize that a nonvanishing $\epsilon_0$ corresponds to a nonvanishing lowest quasiparticle excitation energy $\Delta\epsilon=2\epsilon_0$, and then to a gap in the many-body Hamiltonian spectrum.% and Fig.\ref{fig:evalues_vs_mu_phi_0.4_OBC}(right) at $\Phi/\Phi_0 = 0.4$, for the PBC and OBC cases, respectively. In the PBC case, . \\ 
%------------------------------------------------------------------------------------------------------------------------------------------------------------------------------------------------%
\section{Majorana gap in the clean open chain}\label{app:gap}
\begin{figure}
    \centering
    \begin{tabular}{c}
    (a)\\
     \hspace{0cm}\includegraphics[width=80mm]{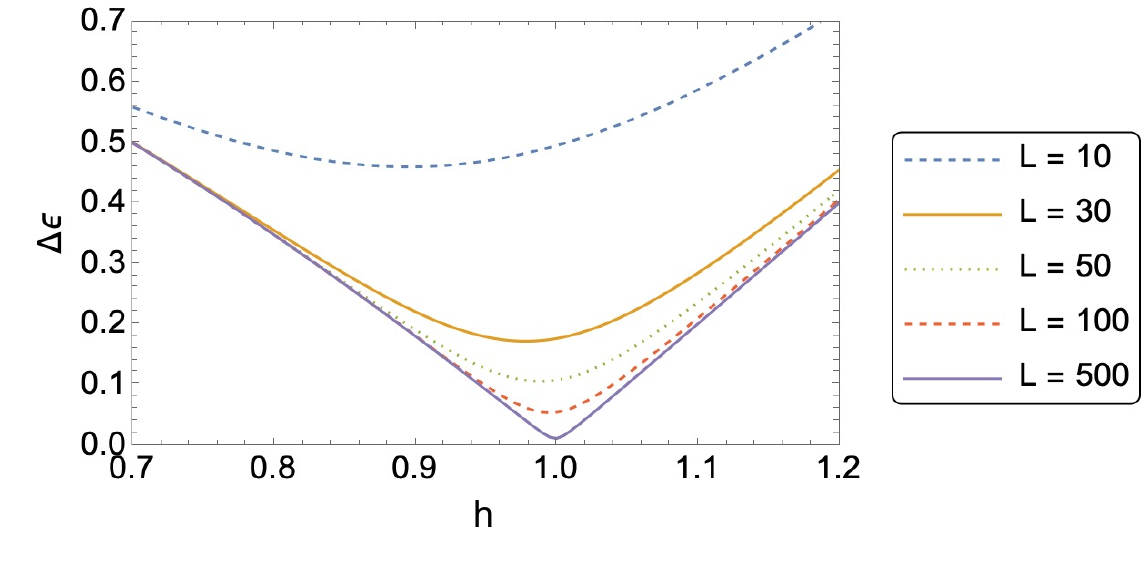}\\
     (b)\\
     \hspace{0cm}\includegraphics[width=80mm]{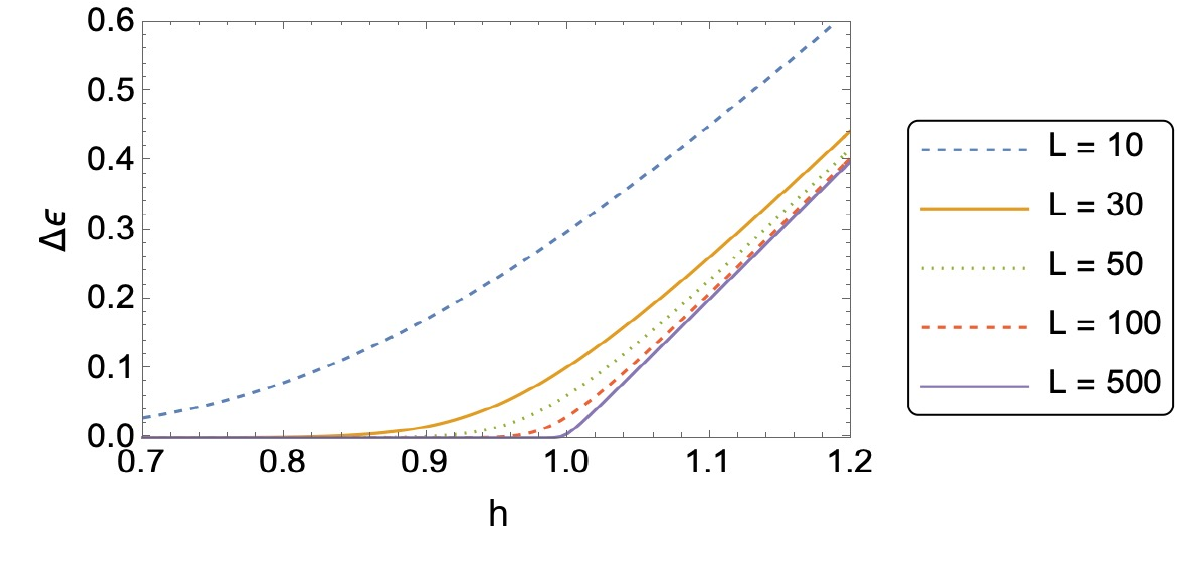}
    \end{tabular}  
    \caption{(Panel a) Lowest excitation energy $\Delta\epsilon=2\epsilon_0$ versus $h$, for $\phi = 0.5$ and different system size $L$ in the clean rf-SQUID. (Panel b) Lowest excitation energy $\Delta\epsilon=2\epsilon_0$ versus $h$, for different system size $L$ in the clean open chain.} 
    %{\bf [Se puoi, al grafico di sotto metti lo 0 sulla linea orizzontale inferiore]}} , at $\phi = 0.4$,
    \label{fig:DEvals_gap}
\end{figure}

{Let fix $\Delta = J = 1$ and consider the rf-SQUID configuration.  A topological gap opens also in the clean case: The lowest excitation energy $\Delta\epsilon=2\epsilon_0$ is nonvanishing in the topological phase $h<J$ [see Fig.~\ref{fig:DEvals_gap}(a)]. Moreover, $\Delta\epsilon$ shows a minimum that becomes smaller and moves towards the critical point with increasing system size, similarly to the disordered case discussed in Sec.~\ref{jj:sec}.}

{In the open chain, instead, a clear mark of the topological phase is the vanishing of the $\Delta\epsilon$ (it is actually exponentially small in $L$). In the trivial phase a finite $\Delta\epsilon$ opens, and remains finite also for large system sizes, at variance with the disordered case [compare with Fig.~\ref{fig:Log(Devals)_vs_h_ALLOPEN}].} %Indeed, considering the lowest excitation energy, one sees that it is vanishing (actually exponentially small in $L$) in the topological phase, and becomes suddenly nonvanishing in the trivial phase, and the transition becomes sharper and sharper and moves towards $h=1$  as the system size increases (see Fig.~\ref{fig:DEvals_gap}).
%
%
%
%\begin{figure}[h!]
%    \centering
%    \begin{tabular}{c}

%    \end{tabular}  
%    \caption{}
%    \label{fig:DEvals_gap_SQUID}
%\end{figure}
%
%
%
%\begin{figure}[h!]
%    \centering
%    \begin{tabular}{c}
%     \hspace{0cm}\includegraphics[width=80mm]{imgs_6_06/Log_DEvals_vs_h_clean_rf_SQUID}
%    \end{tabular}  
%    \caption{Energy gap $\Delta \epsilon$ versus $h$ for the rf-SQUID configuration, for different system size $L$. The logarithmic scale on the vertical axis highlights the minimum that marks the transition between the topological and the trivial phase. Numerical parameters: $\Delta = 1$, $\phi=0.5$.}
%    \label{fig:Log_DEvals_gap_SQUID}
%\end{figure}
%
%
%
%................................................................................................................................................................................................%
\section{Transport through a normal quantum ring}\label{normal:app}
\begin{figure}
    \centering
    (a)\\
    \includegraphics[width=80mm]{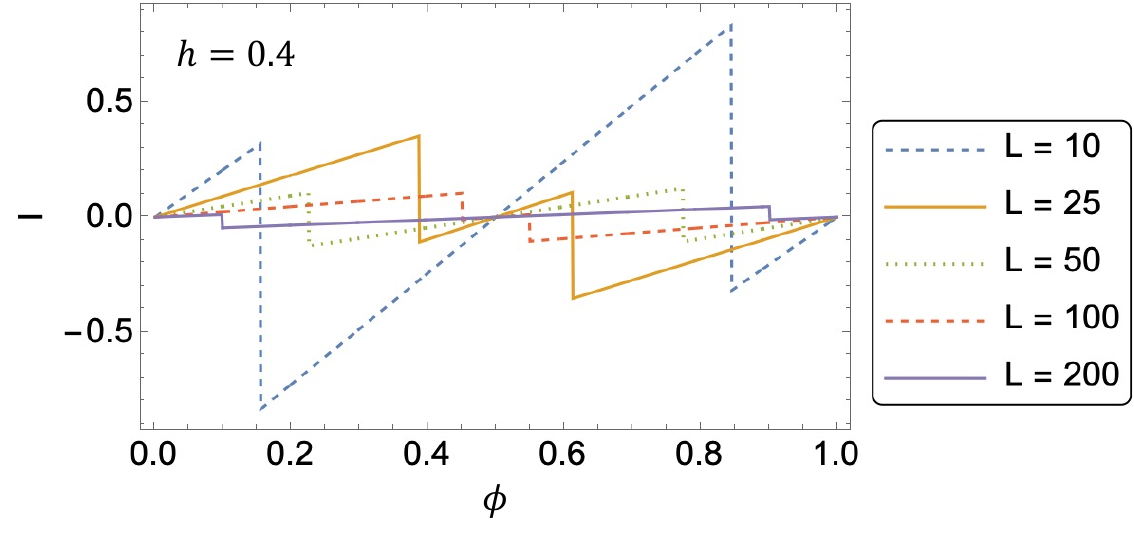}\\
%    (b)\\
%    \includegraphics[width=80mm]{imgs_6_06/Delta_0_I_vs_phi_h_1}\\
%    (c)\\
%         \includegraphics[width=80mm]{I_vs_phi_mu_1.2_Delta_0_L_var}\\
         (b)\\
         \includegraphics[width=80mm]{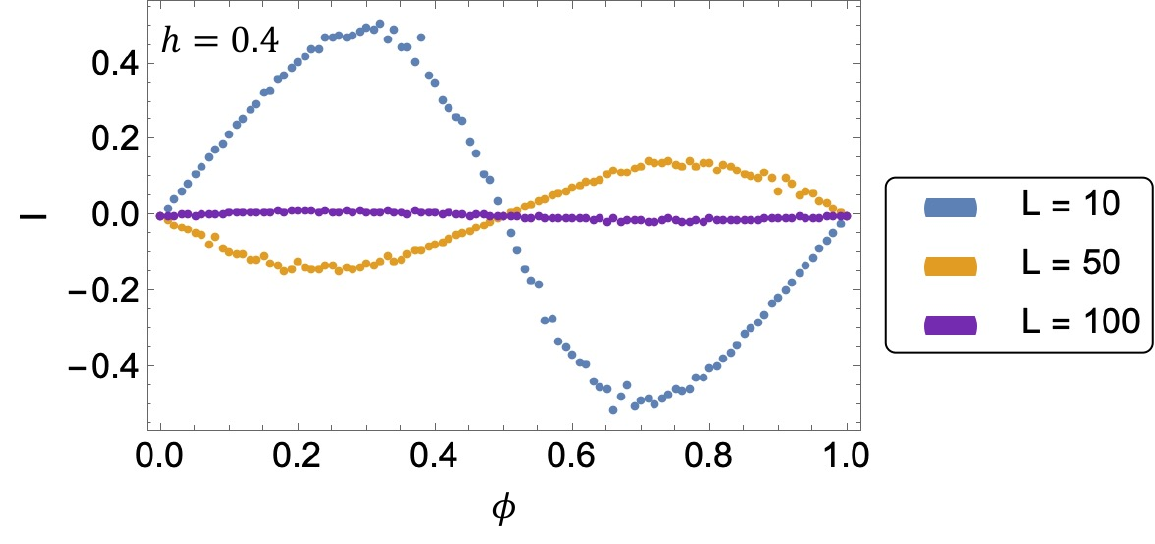}
    \caption{Examples of current  versus $\phi$ for system with no superconducting terms ($\Delta = 0$) and different system sizes $L$. The current always vanishes for large $L$. (a) clean model, $h=0.4$; (b)  dirty model, $h=0.4$.} 
    %{[Metti $\phi$ sull'asse orizzontale.]}} %versus $\Phi/\Phi_0$ at fixed system size $L=50$ in the normal clean sample. (Bottom) Current as a function of $\Phi/\Phi_0$ at $\mu = 0.4$ and $\Delta = 0$, for different system size $L$. (Bottom) Current as a function of $\Phi/\Phi_0$ at $\mu = 1$ and $\Delta = 0$, for different system size $L$. (Bottom) Current as a function of $\Phi/\Phi_0$ at $\mu = 1.2$ and $\Delta = 0$, for different system size $L$. }
    \label{fig:norcle}
\end{figure}

Without the superconducting pairing terms in the Hamiltonian, there is no current in the thermodynamic limit, whatever the value of the chemical potential, either in absence or in presence of disorder (see some examples for the clean case in Fig.~\ref{fig:norcle}). So, the current that we see in presence of the superconducting terms is an effect due to the Cooper-pair condensate.

%..................................................................................%
\section{Some properties of the IPR}\label{IPR:app}
\begin{figure}
    \centering
    (a)\\
    \includegraphics[width=80mm]{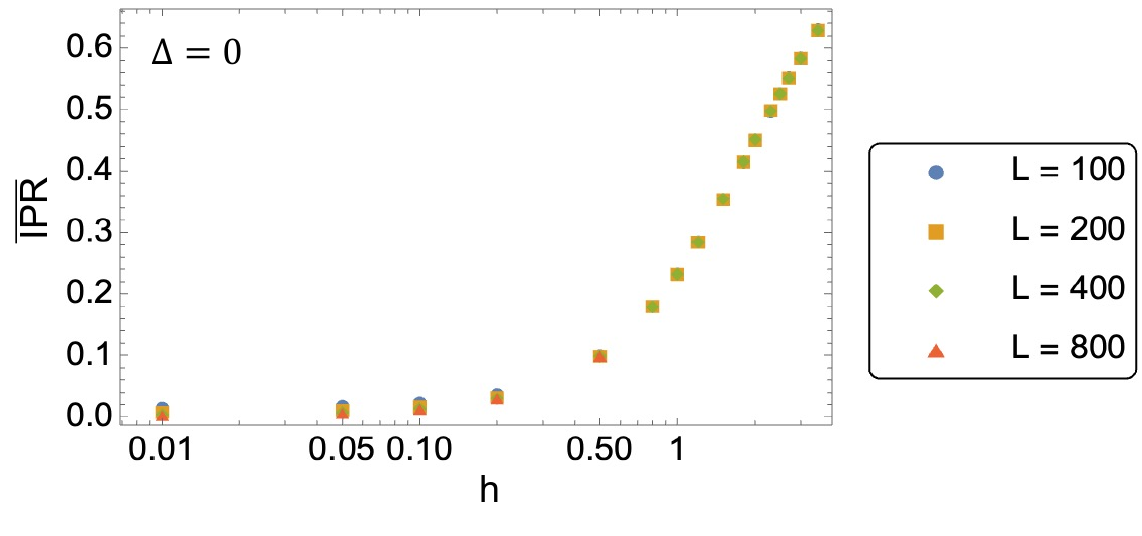}\\
    (b)\\ 
    \includegraphics[width=81mm]{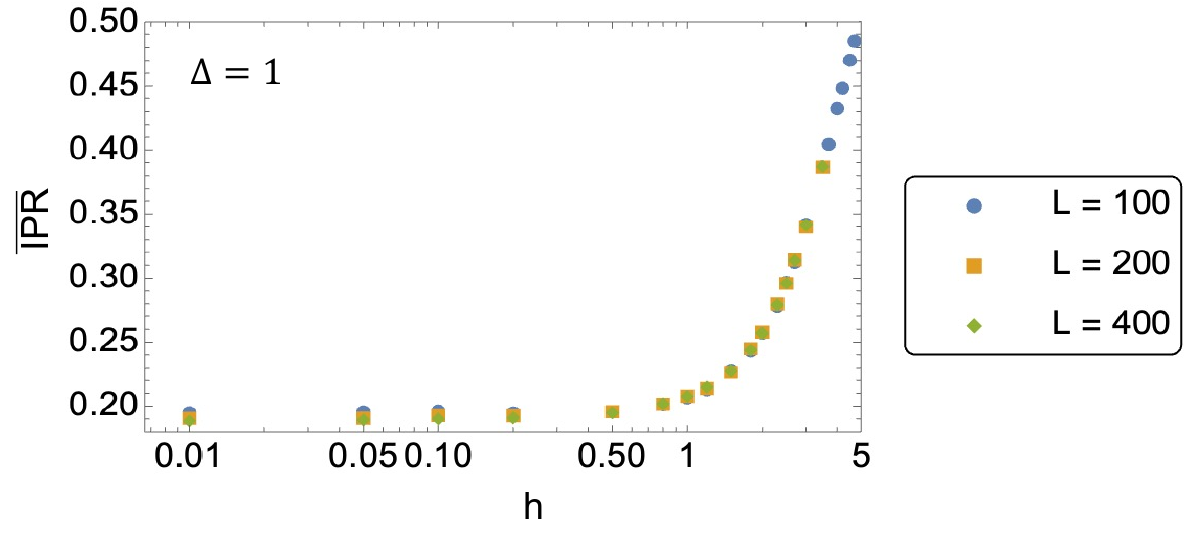}
    \caption{(Panel a) $\overline{\rm IPR}$ versus $h$ in the rf-SQUID, for $\phi = 0.6$, $\Delta = 0$, $N_s=500$. (Panel b) The same for $\Delta = 1$. Notice the logarithmic horizontal axis.}
     %{[Metti anche il caso $\Delta = 1$]}}
    \label{fig:IPR_vs_h_Delta_0}
\end{figure}

{For large disorder strength $h$ the $\overline{\rm IPR}$ defined in Eq.~\eqref{averIPR:eq} scales logarithmically with $h$, both for $\Delta = 0$ (normal case) and $\Delta = 1$ (see Fig.~\ref{fig:IPR_vs_h_Delta_0}). If one considers that the $\overline{\rm IPR}$ is an estimate of the inverse localization length averaged over quasiparticles and disorder, this is in agreement with the theroetical prediction of logarithmic scaling of the inverse localization length with the disorder strength~\cite{scardicchio2017perturbation}.}

\begin{figure}
   \centering
%    (a)\\
     \includegraphics[width=81mm]{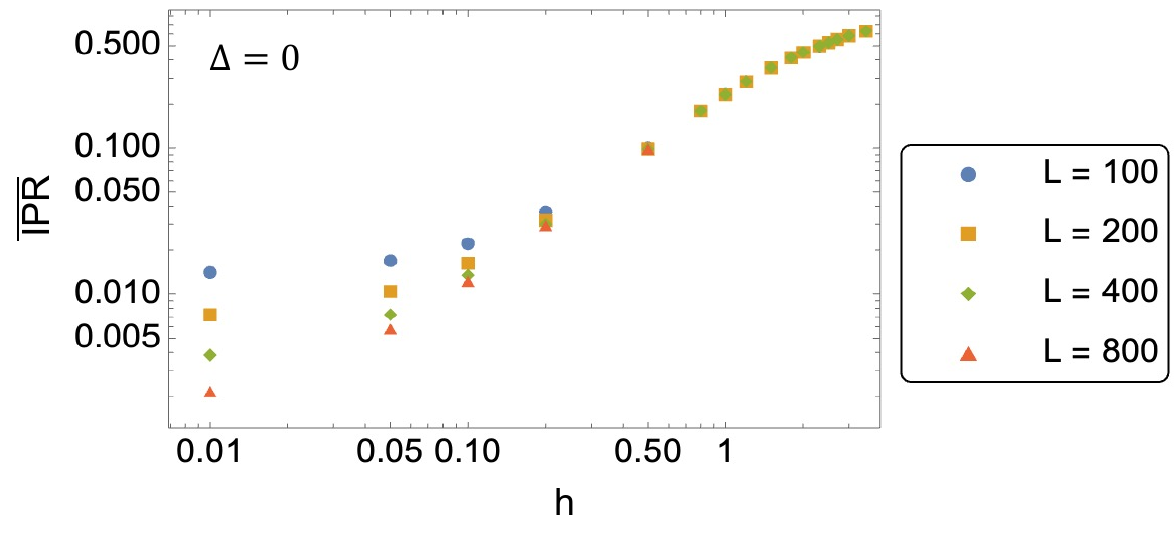}\\
%     (b)\\
%    \includegraphics[width=81mm]{imgs_6_06/LogIPR_vs_Logh_Delta_1} 
    \caption{Bilogarithmic plot of $\overline{\rm IPR}$ versus $h$ in the rf-SQUID, for $\phi = 0.6$, $\Delta = 0$ and $N_{s}=500$.}% In panel (b), the same for $\Delta = 1$.}
    \label{fig:LogIPR_vs_Logh_Delta_0}
\end{figure}

{For small $h$ and $\Delta = 0$, we see that the IPR tends for large system sizes to a power-law behavior (see Fig.~\ref{fig:LogIPR_vs_Logh_Delta_0}). By means of a linear fit of the bilogarithmic plot, we find that the slope of this line (and then the power-law exponent) is $\sim 1.2$. This is smaller than the value $2$ predicted for the inverse localization length of eigenstates at a given energy~\cite{Lif,Kramer_1993,Edwards_1972,Modugno_2010}, due to the fact that in Eq.~\eqref{averIPR:eq} we average over energies.}
%
%
%
%
%
%\bibliography{biblio}

\begin{thebibliography}{78}
\bibitem [{not()}]{nota_errorbar}%
  \BibitemOpen
  \href@noop {} {}\bibinfo {note} {We evaluate the errorbars on the disorder
  average of a quantity as the root mean square deviation divided by
  $\sqrt{N_s}$.}\BibitemShut {Stop}%
\bibitem [{pan()}]{panota}%
  \BibitemOpen
  \href@noop {} {}\bibinfo {note} {In Fig.~\ref{fig:derivate}(b) we see some
  jumps. We have checked that they disappear in the large-size limit, as we
  better discuss in Appendix~\ref{disco:sec}.}\BibitemShut {Stop}%
\end{thebibliography}
\end{document}